\documentclass[preprint,aps,prb]{revtex4-2}
\usepackage{amsmath,mathtools}
\usepackage{amssymb}
\usepackage{gensymb}
\usepackage{amsthm}
\usepackage{graphicx}
\usepackage{physics}
\usepackage {xcolor}
\newcommand{\e}{\varepsilon}
\newcommand{\bea}{\begin{eqnarray}}
\newcommand{\eea}{\end{eqnarray}}
\newcommand{\beq}{\begin{equation}}
\newcommand{\eeq}{\end{equation}}

\begin{document}
	
\title{Stoner Transition at Finite Temperature in a 2D Isotropic Fermi Liquid}

\author{R. David Mayrhofer}
\affiliation{School of Physics and Astronomy, University of Minnesota, Minneapolis, MN 55455, USA}

\author{Megan Schoenzeit}
\affiliation{School of Physics and Astronomy, University of Minnesota, Minneapolis, MN 55455, USA}

\author{Andrey V. Chubukov}
\affiliation{School of Physics and Astronomy and William I. Fine Theoretical Physics Institute, University of Minnesota, Minneapolis, MN 55455, USA}

\date{\today}

\begin{abstract}
We present the results of a mean-field analysis of the temperature evolution of a ferromagnetic Stoner transition in a two-dimensional (2D) system with an isotropic dispersion $\e_k \propto k^{2\alpha}$, which for $\alpha >1$ models flat dispersions in various multi-layer graphene  systems in a displacement field. This study is an extension to a finite $T$ of previous studies at $T=0$,  which found both first-order and second-order Stoner transitions, depending on the value of $\alpha$ and special behavior at $\alpha =1$ and $\alpha =2$. We find that the Stoner transition at a finite $T$ displays new features not seen at $T=0$. The most interesting one is the reentrant behavior, where the ordered state emerges as temperature is increased. This behavior develops at $\alpha >1.4$ and the range where it holds increases with  $\alpha$.  We conjecture that the reentrant behavior is the fundamental feature of the Stoner transition in 2D, not sensitive to the details of the electronic structure.
\end{abstract}
\maketitle

\section{Introduction}
Two-dimensional materials have maintained a high level of interest for the past two decades, beginning with the production of the first flakes of graphene \cite{novoselov2004,novoselov2005,dean2010,radisavljevic2011,geim2013}. In recent years, a plethora of electronically ordered states has been determined in both twisted and untwisted graphene multilayers~\cite{cao2018,cao2018insulator,yankowitz2019,lu2019,andrei2020,cao2021, zhou2021rtg,zhou2021sc,zhou2022bbg,seiler2022,Seiler2024,barrera2022,han2023,han2024,Arp2024,holleis2025}. These ordered states can be classified as spin and/or valley polarized, and the transitions into these states are generally believed to be generalized Stoner transitions~\cite{chichinadze2022,*chichinadze2022letters,haoyu2023,xie2023,lee2024,koh2024rtg,koh2024bbg,wang2024electrical,friedlan2025,mayrhofer2025}.

A ferromagnetic Stoner transition has also been identified within a generic Fermi liquid theory as a Pomeranchuk instability, developing when the Landau parameter $F^s_0 =-1$, where $s$ stands for spin and $0$ for  the channel with zero angular momentum.
A Stoner transition in 3D has been studied within mean-field (ladder) approximation both at $T=0$ (Ref. \cite{stoner1938}) and at a finite $T$ (Ref. \cite{Shimizu1981}). The transition is continuous (second-order) both at $T = 0$ and $T>0$. At $T=0$ the transition occurs when the dimensionless coupling $\lambda$ (the product of the interaction $g$  and the density of states at the Fermi level $\nu_F$) reaches $\lambda =1$.  At a finite $T$, the transition temperature $T_c$ increases with $\lambda$, consistent with a common-sense expectation that thermal fluctuations act against long-range order.
A 3D Stoner transition has also been of interest in the context of atomic Fermi gases \cite{duine2005,he2016}.

In 2D, a Stoner transition for fermions with an isotropic dispersion has been recently analyzed at $T=0$ both analytically and numerically \cite{zhu2018,valenti2024,calvera2024nematicity,raines2024stoner}, and the results have been applied to untwisted multi-layer graphene sheets in a displacement field and also to AlAs quantum wells, which is another two valley system which exhibits valley and spin polarization
 ~\cite{shayegan2006,gunawan2006,hossain2020,hossain2021,hossain2022}. Analytical studies within the mean-field (ladder) approximation~\cite{raines2024unconventional,*raines2024isospin} have found that the transition is rather unconventional, particularly for $k^2$ and $k^4$ dispersions. Namely, the order parameter susceptibility diverges when the coupling $\lambda \to 1$, as is expected for a continuous, second order transition. However, immediately below the transition the order parameter instantly jumps to the largest possible value. For a single-valley system, this leads to a ferromagnetic half-metal. For a two-valley system that allows both  valley and spin orders, the ordered states are either half-metal with spin or valley order and quarter-metal with spin and valley order.  Both half-metal and quarter-metal ordered states have been obtained~\cite{valenti2024} in the variational Quantum Monte Carlo analysis of a two-valley system with  Coulomb interaction and anisotropic $k^2$ dispersion, applicable to AlAs.

In this communication  we extend the mean-field analysis of Stoner transition in 2D to finite temperature. To the best of our knowledge, no such study has been performed so far.  We consider a ferromagnetic transition in a one-valley system with interacting spin-half fermions with a dispersion $\e_k \propto k^{2\alpha}$, where $\alpha>0$.  An extension to a two-valley system is straightforward and can be done in the same way as in Refs. [\onlinecite{raines2024unconventional,*raines2024isospin}].

At $T=0$, it was determined that Stoner transition is continuous second-order for $\alpha <1$ and $\alpha >2$,
 and first-order into a fully polarized half-metal state for $1 \leq \alpha \leq 2$, with special behavior for $\alpha =1$
 ($\epsilon_k \propto k^2$) and $\alpha =2$ ($ \epsilon_k \propto k^4$) that we already mentioned. A first-order transition occurs at $\lambda = 2(2^\alpha-1)/(\alpha (\alpha +1))$, which for all $1\leq \alpha \leq 2$ is very close to $\lambda =1$, where the order parameter susceptibility would diverge if the transition was a continuous one.

At finite temperature, we find that the results  depend even more strongly on the value of $\alpha$. For $\alpha<1$,  the transition remains second order and $T_c (\lambda)$
 increases with the coupling $\lambda$ (see Fig. \ref{linear}). This is the same behavior as in 3D. In the special case $\alpha =1$, we reproduce singular behavior of the magnetization at $T \to 0$, but at a finite $T$ the magnetization evolves continuously below $T_c(\lambda) = \e_F/\log \left(\lambda/\left(\lambda -1 \right)\right)$ as a function of either $T_c-T$ or $\lambda - \lambda_c (T)$. We show this behavior in Fig. \ref{quadratic}.
 For $1<\alpha<2$, we find that  at small $T$ the transition becomes first order, like at $T=0$, but now it is into a partly polarized state. The jump of the order parameter at the transition shrinks with increasing $T$, and above a certain temperature Stoner transition becomes second-order. This is an expected behavior given that the transition is first-order at $T=0$ and that it  is continuous at $T >0$  at the boundary value $\alpha =1$.

 For $\alpha <1.4$, $T_c$ for both first-order and second-order transition increases with $\lambda$, i.e., the topology of the phase diagram is the same as at $\alpha <1$.  However, for $\alpha >1.4$, the topology changes and a reentrant behavior emerges, first near a particular $T_c (\lambda)$, where a first order transition becomes second-order and then in a finite range along $T_c (\lambda)$ line, including a range where the transition is second-order.  We show this behavior in Fig. \ref{reentrant}.
As $\alpha$ increases towards $2$, the upper and lower bounds of the range of reentrant behavior evolve differently. The upper boundary, where the transition is second-order, remains at a finite $T_c$, but the lower boundary, where the transition is first-order, extends down towards $T=0$ and the jump of the order parameter at the lower end of first-order transition range transition gets larger.  At $\alpha = 2$, the reentrant behavior extends down to $T=0$ (see Fig. \ref{quartic}).  At this $\alpha$, the dependence of $T_c (\lambda)$ is non-analytic at small $T$: $T_c (\lambda) \propto (1-\lambda)^{2/3}$.

At $\alpha \geq 2$, the transition at the lowest $T$ becomes reentrant second order and the range of $T_c (\lambda)$  where the transition remains reentrant first order gets sandwiched between second-order transitions at both lower and higher $T$.  As $\alpha$ increases, this range shrinks and disappears at $\alpha \approx 2.15$. This progression is shown in Fig. \ref{alpha>2}. At larger $\alpha$, the transition is second order for all $T$, reentrant at smaller $T$ and non-reentrant at larger $T$. As $\alpha$ increases further, the topology of the phase diagram remains unchanged, but $T_c (\lambda)$ curve displays a finger-type extension towards smaller $\lambda$. In the limit $\alpha \to \infty$, when the dispersion at small $k$ is almost completely flat, the Stoner instability near the tip of $T_c (\lambda)$ curve develops already at infinitesimally small $\lambda \sim 1/\log\alpha$.

Some of the fine features that we find may be specific to the particular form of the dispersion that we are using.  However the reentrant behavior is a robust feature of our phase diagram for sufficiently flat fermionic dispersion. Given that a generic tight-binding model for a multi-layer untwisted graphene yields a dispersion that gets flatter with the increase of the number of layers \cite{zhang2010,min2008} and that quantum oscillations and compressibility experiments on these materials  detected sharp transitions into half-metal and quarter-metal phases upon variation of either number density or displacement field~\cite{seiler2022,zhou2022bbg, zhou2021rtg,Arp2024,holleis2025,zhang2023,zhou2021sc,Patterson2024}, we expect that our results about reentrant behavior are applicable to multi-layer graphene sheets and call for measurements of the thermal evolution of e.g., critical displacement field for a transition into a half-metal at a fixed density ~\footnote{There is an evidence that a half-metal state is spin-polarized~\cite{barrera2022,seiler2022}.}.  We also note in passing that reentrant behavior of $T_c$ has been analyzed in the context of symmetry breaking in conformal field theories~\cite{Chai2020}.

The paper is organized as follows. In Sec. \ref{mean_field}, we present the Hamiltonian for the system alongside the mean field formalism used to solve for the order parameter. In Sec. \ref{ordered_states}, we present both the analytic and numerical solutions to the order parameter, the chemical potential and the free energy as functions of temperature and the coupling and obtain the phase diagrams for different $\alpha$. In Sec. \ref{conclusion}, we present our conclusions. Some technical calculations are moved into Appendices.

\section{Mean-Field Equations}
\label{mean_field}
Our point of departure is a Hamiltonian for spin 1/2 fermions within one valley with the dispersion $\e_k$ and short range interaction $g$:
\begin{align}
H = \sum_{\vb k, \alpha} \e_{\vb k} c^{\dagger}_{\vb k, \alpha} c_{\vb k, \alpha} + \frac{g}{2V}\sum_{\vb k, \vb p, \vb q,\alpha,\beta} c^{\dagger}_{\vb k + \vb q, \alpha} c^\dagger_{\vb p - \vb q, \beta} c_{\vb p, \beta} c_{\vb k, \alpha}.
\end{align}
We assume that the dispersion is isotropic and takes the form,
\begin{align}
\e_{k} = c \left(\frac{k^2}{4\pi}\right)^{\alpha}.
\end{align}
The corresponding density of states  at the Fermi level $\nu_{F,\alpha}$  is
\begin{align}
\nu_{F,\alpha} = \frac{\e_F^{\frac{1}{\alpha} - 1}}{\alpha c^{1/\alpha} }.
\end{align}
The dimensionless coupling  $\lambda$ is the product of $g$ and the density of states:
\begin{align}
\lambda = g \nu_{F, \alpha}.
\end{align}

As we said, we will analyze the type of transition into a ferromagnetic state and the topology of the phase diagram within the mean-field (ladder) approximation. It has been known for quite some time~\cite{Kanamori1963} that this approximation under-estimates the strength of fluctuations which tend to eliminate a Stoner instability or at least move it to larger $\lambda$. On the other hand,  the results of the $T=0$ mean-field analysis on the development of  spin and valley  polarization in 2D are largely consistent with the data for multi-layer graphene sheets (Bernal bilayer graphene, rhombohedral trilayer graphene, rhombohedral pentalayer graphene) and AlAs (see more on this in Ref. \cite{raines2025}).
We take this as a justification to extend mean-field analysis to finite $T$.

Within mean-field, the ferromagnetic order parameter is determined self-consistently as
\begin{align}
\label{self-consistent} \Delta = \frac{g}{2V} \sum_{\vb k, \alpha,\beta} \langle c^{\dagger}_{\vb k,\alpha} \sigma_{\alpha \beta}^{z} c_{\vb k, \beta} \rangle= \frac{g}{2V} \left(N_{\uparrow} - N_{\downarrow}\right),
\end{align}
where $V$ is the area of the system.
This self-consistent equation relates the spin order parameter to spin dependent energy shift of the bands in the ordered state, induced by the same order parameter. The system must also obey particle conservation, which is the condition that the number of particles,
\begin{align}
\label{particle} N = \sum_{\vb k, \alpha,\beta} \langle c^{\dagger}_{\vb k,\alpha} \delta_{\alpha \beta} c_{\vb k, \beta} \rangle,
\end{align}
 or, equivalently, particle density $n = N/V$ does not change upon spin polarization. This equation determines the temperature dependent chemical potential, which in the normal state we call $\mu (T)$ and in the ordered state ${\bar \mu} (T)$. The particle density can be easily evaluated and equals
\begin{align}
n = 2 \alpha \epsilon_F \nu_{F, \alpha}.
\label{s_1}
\end{align}
Using (\ref{self-consistent}) and (\ref{particle}), we can relate $\Delta$  to the polarization of the system, $\zeta = (N_{\uparrow} - N_{\downarrow})/N$:
\begin{align}
\Delta = \zeta \alpha \lambda \e_F.
\label{s_2}
\end{align}
Lastly, we note the true ground state is a minimum of the free energy of the system. If the system undergoes a second order phase transition, it is essentially guaranteed that once the self-consistent ladder equation  (\ref{self-consistent}) develops a non-zero solution, this solution minimizes the free energy.  However, if the transition is first order,
one must explicitly compare the free energies of the system at finite $\Delta$ and at $\Delta=0$.
To do this, we need the expression for the free energy.  Within the ladder approximation, it is
\begin{align}
\label{energy} F_{ord} = \Omega_{ord} + \bar{\mu} N = \frac{V \Delta^2}{g} + \bar \mu N - T \sum_{k,\sigma} \log \left(1 + e^{\left(\bar \mu - \e_{k, \sigma}\right)/T} \right),
\end{align}
where $\Omega_{ord}$ is the grand potential in the ordered state and $\e_{k, \sigma} = \e_{k} \pm \Delta$.  The free energy in the normal state is obtained from Eq. (\ref{energy}) by setting $\Delta = 0$ and ${\bar \mu} = \mu$.

\section{Ordered States}
\label{ordered_states}
Here, we obtain the order parameter, the chemical potential, and the free energy at different values of $\alpha$ and detect the phase boundary between the ordered and normal states.

First, assume that the transition is second order. The order parameter susceptibility then diverges precisely at the transition point. Within the ladder approximation, the susceptibility in the normal state is
\begin{align}
\label{susceptibility} \chi = \frac{\chi_0 }{1-g \Pi_{\alpha}(T)},
\end{align}
where $\Pi_{\alpha}(T)$ is the static particle-hole bubble at zero-momentum transfer. The susceptibility diverges at  $g\Pi_{\alpha}(T) = 1$. The polarization $\Pi_\alpha(T)$ at a finite temperature can be obtained analytically:
\begin{align}
\nonumber \Pi_\alpha(T) &= -T \sum_{n} \int \frac{d^2k}{\left(2\pi \right)^2} G(\vb k, \omega_n)^2 = - \nu_{F,\alpha} \e_F^{1-\frac{1}{\alpha}}\int \limits_{0}^{\infty} d\e_k \e_k^{\frac{1}{\alpha}-1} n_F'(\e_k-\mu) \nonumber \\
 &= - \nu_{F,\alpha}\Gamma \left(\frac{1}{\alpha}\right) \left(\frac{T}{\e_F}\right)^{\frac{1}{\alpha} -1} \text{Li}_{\frac{1}{\alpha}-1}\left(-e^{\mu/T}\right),
 \label{polarization_eq}
\end{align}
where $G(\vb k, \omega_n) = (i \omega_n -\e_k-\mu)^{-1}$ is the fermionic Green's function, and Li$_{s}(z)$ is the polylogarithm (the Jonquiere's function). Similarly, the order parameter $\Delta$, the chemical potential ${\bar \mu}$ and the free energy are also expressed via polylogarithms:
\begin{align}
\label{eq1} \Delta &=
\frac{g}{2V} \sum_{\beta} \int \frac{d^2k}{(2\pi)^2} n_F(\e_{k,\beta}-\bar \mu)\sigma_{\beta \beta}^{z}
= \frac{\lambda}{2} \e_F^{1-\frac{1}{\alpha}} \sum_{\beta} \int \limits_{0}^{\infty} d\e_k \e_k^{\frac{1}{\alpha}-1} n_F(\e_{k,\beta}-\bar \mu)\sigma_{\beta \beta}^{z} \\
\nonumber &= -\frac{\lambda \e_F}{2} \left(\frac{T}{\e_F} \right)^{\frac{1}{\alpha}} \Gamma \left(\frac{1}{\alpha}\right) \left[ \text{Li}_{\frac{1}{\alpha}} \left(-e^{\left(\bar \mu + \Delta \right)/T} \right) -\text{Li}_{\frac{1}{\alpha}} \left(-e^{\left(\bar \mu - \Delta \right)/T} \right) \right], \\
\label{particle_number} n &= 2\alpha \nu_{F,\alpha} \e_F =
- \nu_{F,\alpha}\e_F \left(\frac{T}{\e_F}\right)^{\frac{1}{\alpha}} \Gamma\left(\frac{1}{\alpha}\right)
\left[ \text{Li}_{\frac{1}{\alpha}} \left(-e^{\left(\bar \mu + \Delta \right)/T} \right)
+\text{Li}_{\frac{1}{\alpha}} \left(-e^{\left(\bar \mu - \Delta \right)/T} \right) \right],\\
\label{freeenergy} \frac{F_{ord}}{N} & =
\frac{\Delta^2}{2 \alpha \e_F \lambda } + \bar \mu +
\frac{\e_F}{2\alpha} \left(\frac{T}{\e_F}\right)^{\frac{1}{\alpha} + 1}
\Gamma\left(\frac{1}{\alpha}\right)
\left[\text{Li}_{1+\frac{1}{\alpha}} \left(-e^{\left(\bar \mu + \Delta \right)/T} \right)
+\text{Li}_{1+\frac{1}{\alpha}} \left(-e^{\left(\bar \mu - \Delta \right)/T} \right) \right].
\end{align}

\begin{figure}[h]
	\begin{center}
		\includegraphics[scale=.8]{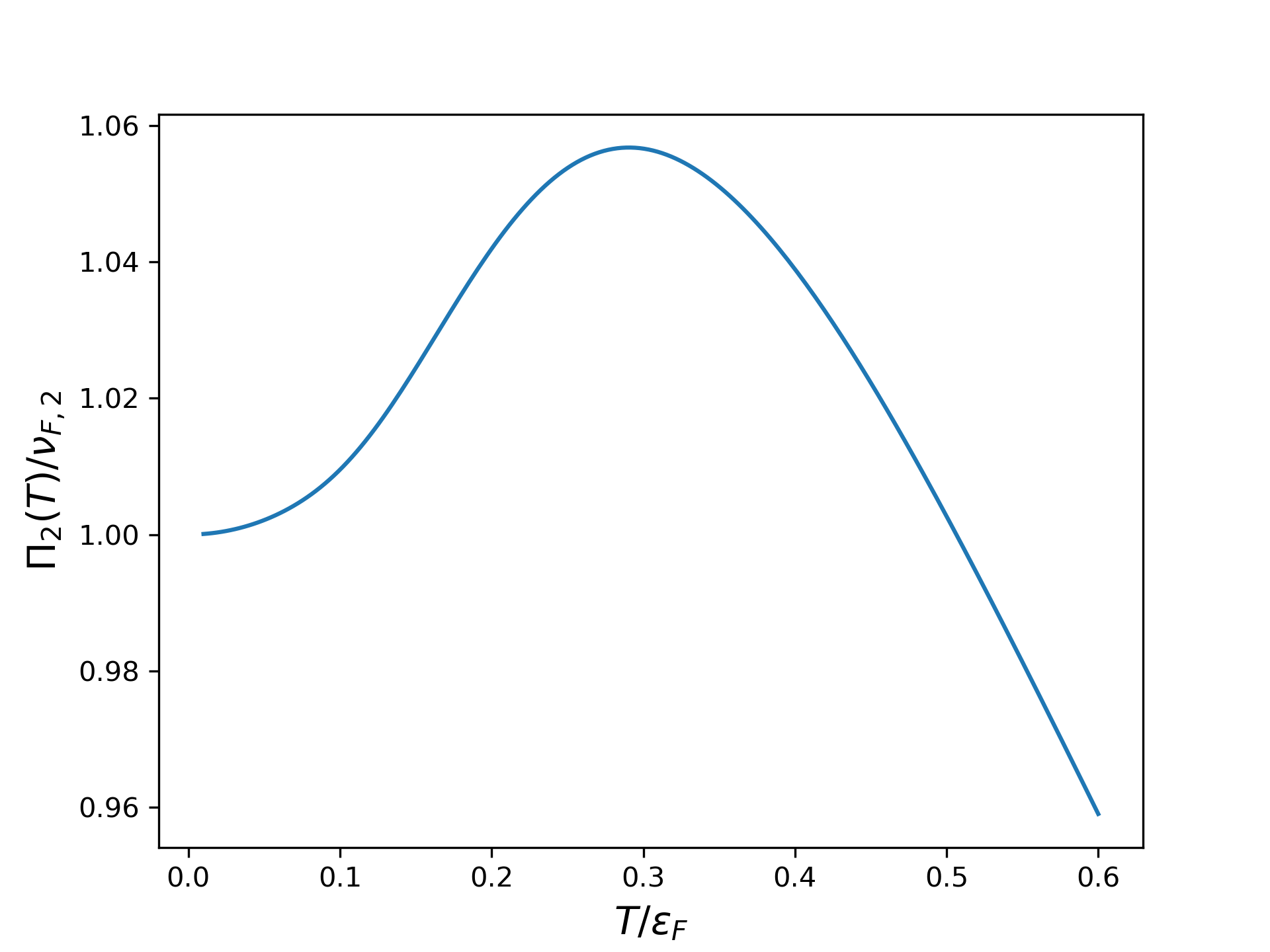}
		\caption{The polarization bubble plotted as a function of temperature for $\alpha=2$. One can see that the polarization bubble increases at low temperatures before decreasing at large temperatures.}
		\label{polarization}
	\end{center}
\end{figure}

One can extract the low temperature behavior of the polarization bubble and other variables by expanding the polylogarithm in powers of temperature. For $z \gg 1$,
\begin{align}
\label{polylog} \text{Li}_s\left(-e^{z} \right) =  -2 \sum_{k=0}^{\infty} \eta(2k) \frac{z^{s-2k}}{\Gamma \left(s-2k+1\right)} + \mathcal{O}\left(e^{-z}\right),\\
\eta(k) = -2\pi^{k-1} \sin\left(\frac{\pi k}{2}\right) \left(1-2^{k-1}\right)\Gamma(1-k)\zeta(1-k),
\end{align}
where $\zeta(k)$ is the Riemann zeta function \cite{polylog} and  $\eta(0) = 1/2$, $\eta(2) = \pi^2/12$. The $k=0$ term corresponds to the $T=0$ contribution, and the $k=1$ term gives the $T^2$ correction.  For the chemical potential in the normal state this gives  (see Appendix A for details)
\begin{align}
\label{norm_chem} \frac{\mu(T)}{\e_F} = 1+\frac{\pi^2}{6} \frac{\left(\alpha -1\right)}{\alpha}\left(\frac{T}{\e_F}\right)^2 + \mathcal{O}\left(T^4\right).
\end{align}
Inserting this into Eq. (\ref{polarization_eq}), we find a similar looking formula
\begin{align}
\label{susceptibility_t2} \Pi_\alpha(T) = \nu_{F,\alpha} \left(1 + \frac{\pi^2}{6} \frac{\left(\alpha -1\right)}{\alpha}\left(\frac{T}{\e_F}\right)^2 + \mathcal{O}\left(T^4\right) \right).
\end{align}
Note that for $\alpha>1$, the coefficient for the $T^2$ term is positive. This indicates that the susceptibility in Eq. (\ref{susceptibility}) diverges at a smaller $\lambda$ at finite temperature than at $T=0$, signaling the possibility of reentrant behavior. We plot the full polarization for $\alpha = 2$ in Fig. \ref{polarization} for arbitrary $T/E_F$ using Eq. (\ref{polarization_eq}) with $\mu$ extracted numerically from Eq. (\ref{particle_number}) at $\Delta =0$.
 We will consider three separate cases: $0< \alpha \leq 1$, $1<\alpha<2$, and $2 \leq \alpha$.

\subsection{$0< \alpha \leq 1$}
At $T=0$, it has already been established that the transition is second order, i.e., the order parameter evolves continuously at the transition point, where the susceptibility diverges. At $\alpha=1$, the phase transition is unconventional: the  susceptibility diverges at the transition yet the order parameter jumps discontinuously to its maximum value.

\begin{figure}[h]
	\begin{center}
		\includegraphics[scale=.8]{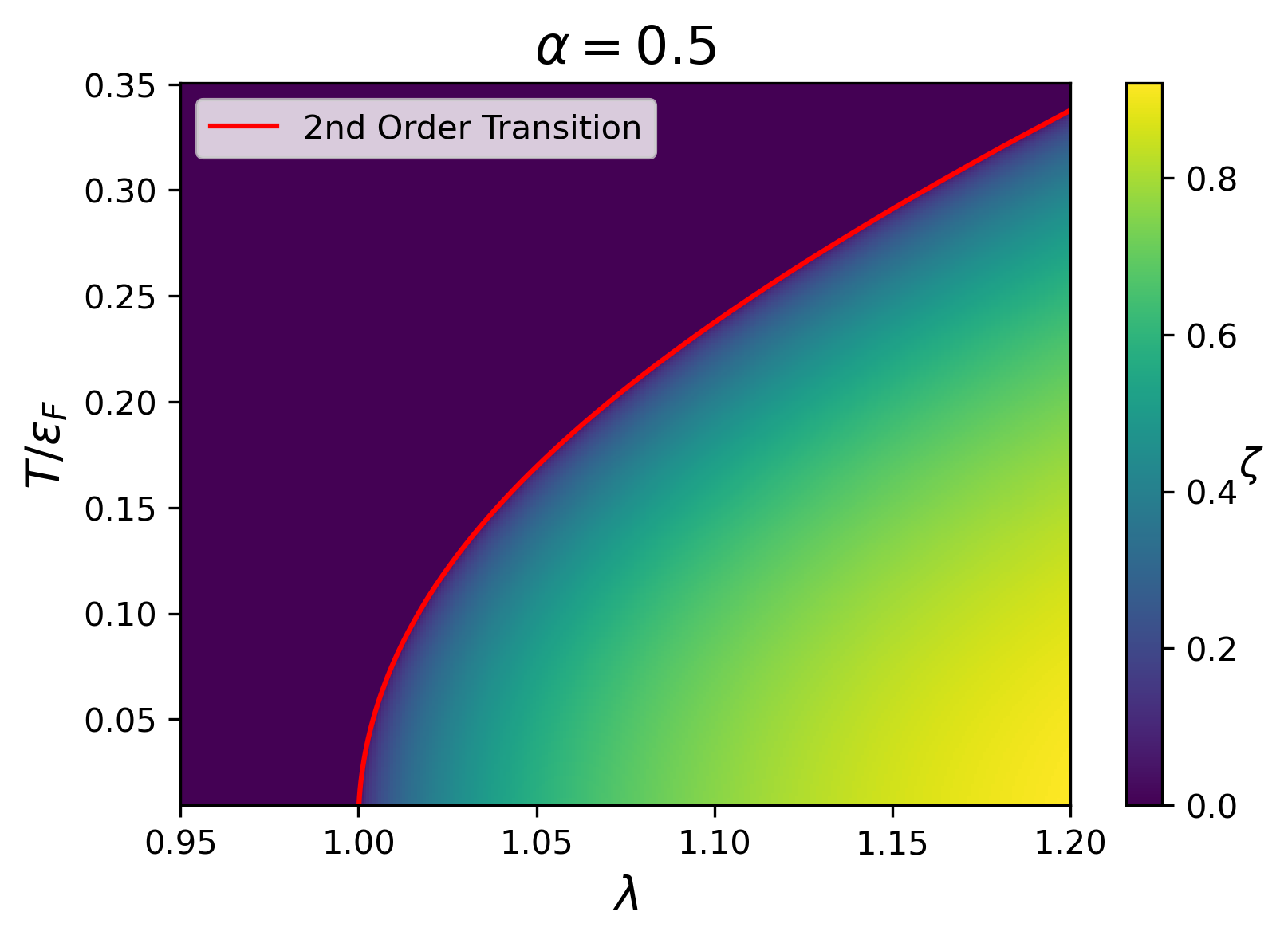}
		\caption{The polarization $\zeta$ when $\alpha =1/2$, i.e. when $\e_k \propto |k|$. The transition is second order for all $T$. The red line is where $\chi$ diverges. }
		\label{linear}
	\end{center}
\end{figure}

At finite temperature, we find that the transition remains second order. The order parameter evolves continuously throughout the phase diagram, as seen in Fig. \ref{linear}, where we plot the polarization $\zeta$ for $\alpha =1/2$.

\begin{figure}[h]
	\begin{center}
		\includegraphics[scale=.8]{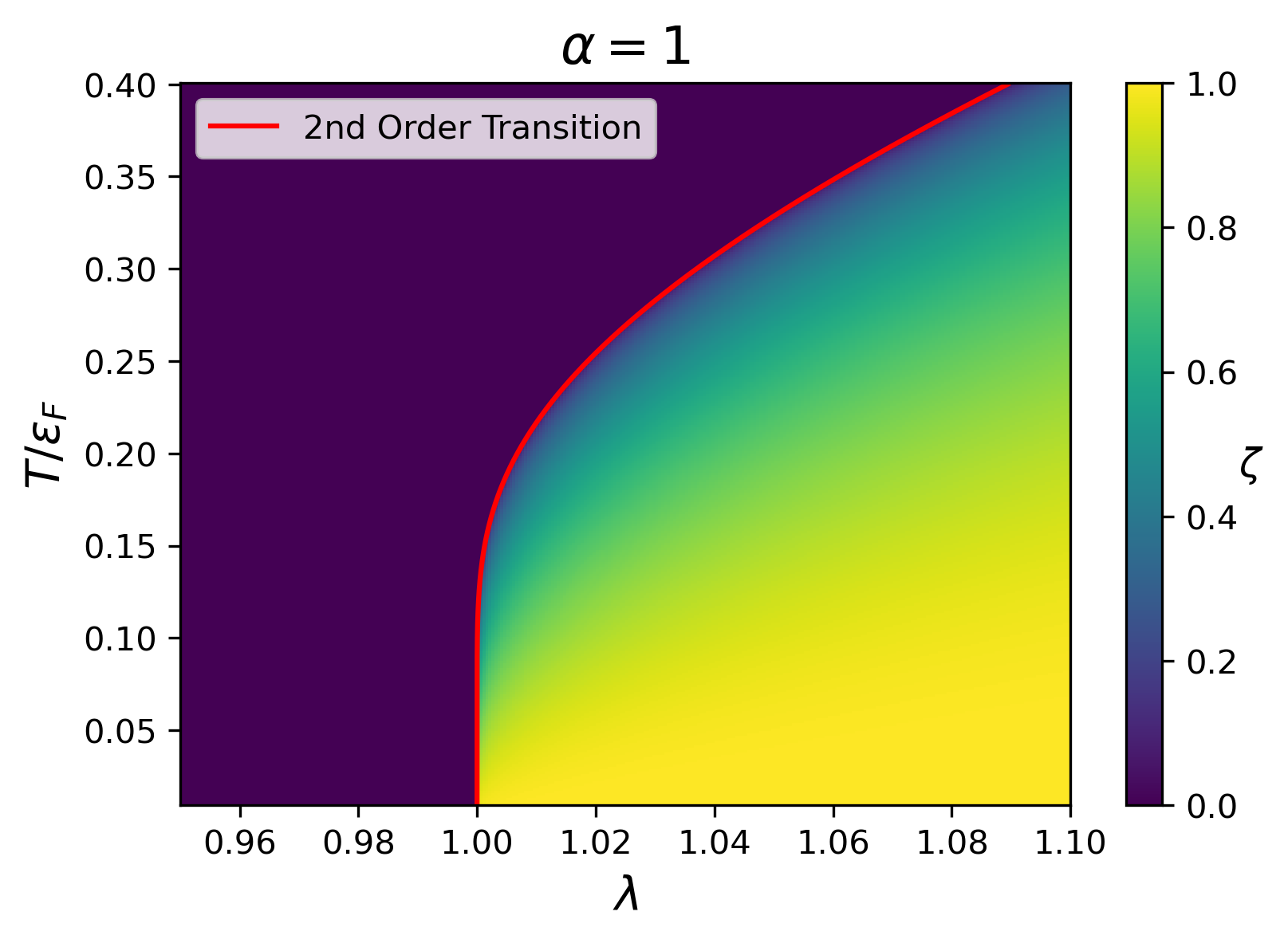}
		\caption{The polarization $\zeta$ when $\alpha =1$. The transition throughout the phase diagram is second order, with the exception of $T=0$, where the order parameter changes discontinuously.}
		\label{quadratic}
	\end{center}
\end{figure}

For $\alpha=1$, we find that at any finite temperature the order parameter smoothly increases with increasing $\lambda$.  For this $\alpha$, the chemical potential in the normal state  and the particle-hole bubble can be evaluated exactly:
\begin{align}
\mu(T) &= T \log \left(e^{\frac{\e_F}{T}} -1 \right),
\\
\Pi_{1}(T) &= \nu_{F,1} \left( 1 - e^{-\frac{\e_F}{T}} \right).
\end{align}
The condition for the system to go through a second order phase transition is $g \Pi_1 (T)$, i.e.,  
\begin{align}
\lambda_c(T) = \frac{1}{1-e^{-\frac{\e_F}{T}}}.
\label{q_1}
\end{align}
We plot this $\lambda_c(T)$ alongside the numerical solution to the self-consistent equation for $\Delta$ in Fig. \ref{quadratic}. We see that the range of $T$ where the polarization $\zeta$ changes gradually shrinks as temperature decreases. At $T=0$, one recovers the result that the system is fully polarized immediately below the onset of order.

\begin{figure}[h]
	\begin{center}
		\includegraphics[scale=.6]{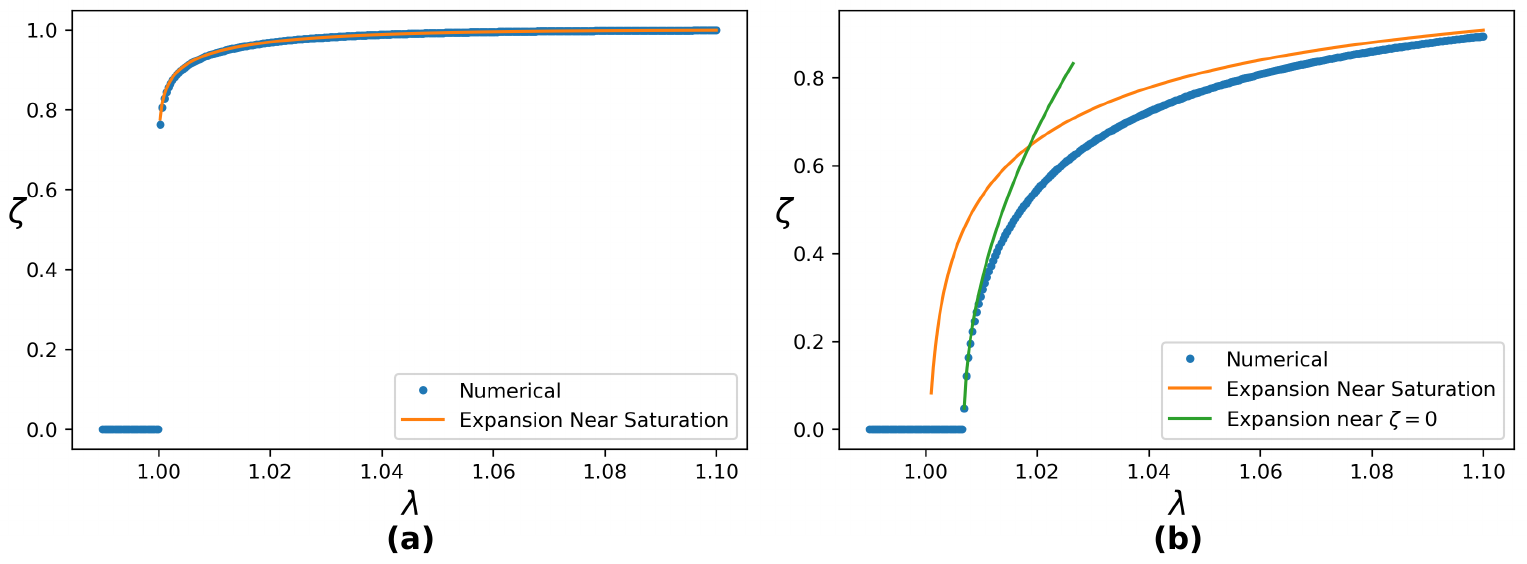}
		\caption{The polarization at fixed temperature at $\alpha=1$ (a) $T=0.05\e_F$ and (b) $T=0.2\e_F$. Blue line - the numerical result,  green and orange lines -- the analytic expressions,  Eqs.(\ref{small_quadratic_analytic}) and (\ref{quadratic_analytic})  respectively. }
		\label{quadratic_polarization}
	\end{center}
\end{figure}

To confirm this is the case, we analytically calculate the order parameter for $\alpha = 1$.
We consider the order parameter in two different limits. First, we consider $\Delta$ in the immediate vicinity of $\lambda_c (T)$,  where $\Delta$ is small. In this case, one can expand the self-consistent equation in powers of $\Delta$ and solve as a function of $T$ and $\lambda$. We present the details of the calculation in Appendix \ref{alpha=1}.
The result is
\begin{align}
\label{small_quadratic_analytic} \Delta = T e^{\frac{\e_F}{2T}}\sqrt{6\left( \lambda - \lambda_c(T)\right)}.
\end{align}
We see here that the order parameter displays a conventional mean-field behavior, $\Delta \propto \sqrt{\lambda - \lambda_c(T)}$ where $\lambda_c (T)$ is given by (\ref{q_1}).
We then analyze the behavior of the order parameter close to its saturated value  at $\lambda >1$. We again present the calculation in Appendix \ref{alpha=1}, and write the result here,
\begin{align}
\label{quadratic_analytic} \Delta  =  \lambda \left(\e_F + T \log \left( 1- e^{-\frac{2\e_F\left(\lambda -1 \right)}{T}}\right)\right).
\end{align}
At $T \to 0$, the second term in (\ref{quadratic_analytic}) goes to zero and $\Delta$ jumps to $\e_F$ already at $\lambda =1$. However, at finite temperature, $\Delta$ continuously approaches the saturation value as $\lambda$ increases. This is consistent with the numerical results. The expansion in (\ref{quadratic_analytic}) breaks down close to $\lambda -1 \sim \frac{T}{2\e_F} |\log \left( 1- e^{-\frac{\e_F}{T}} \right)|$. In Fig. \ref{quadratic_polarization} we compare this expansion to the numerically evaluated order parameter. At low temperatures, the analytic expression matches very closely with the numerics.

\subsection{$1<\alpha<2$}
For these $\alpha$, the  $T=0$ transition is first order and it occurs before the susceptibility diverges. To determine where the phase transition occurs, we directly analyze the free energy of the system.

We present the details of the calculation in Appendix \ref{alpha>1} and here list the results. We find
\begin{align}
\frac{\Delta}{\e_F} &= \alpha \lambda + \mathcal{O}\left( e^{\frac{\bar \mu - \alpha \lambda}{T}}\right),\\
\frac{\bar \mu(T)}{\e_F} &= 2^{\alpha} - \alpha \lambda +  \frac{\left(\alpha-1\right) \pi^2}{3 \alpha}\frac{1}{2^{1+\alpha}} \left( \frac{T}{\e_F}\right)^2,\\
\label{lambda1} \lambda_c(T) &= \frac{2\left(2^\alpha-1\right)}{\alpha \left(\alpha+1 \right)} + \frac{\left(1-2^{-\alpha}\right)}{\alpha^2} \pi^2 \left(\frac{T}{\e_F}\right)^2.
\end{align}
The expansion holds when ${\bar \mu} (T) < \Delta (T)$. This condition is satisfied at $T=0$ and, by continuity, at small finite $T$.
Close to $\alpha = 1$ and $\alpha =2$, the range of temperatures for which this condition is applicable shrinks.

\begin{figure}[h]
	\begin{center}
		\includegraphics[scale=.44]{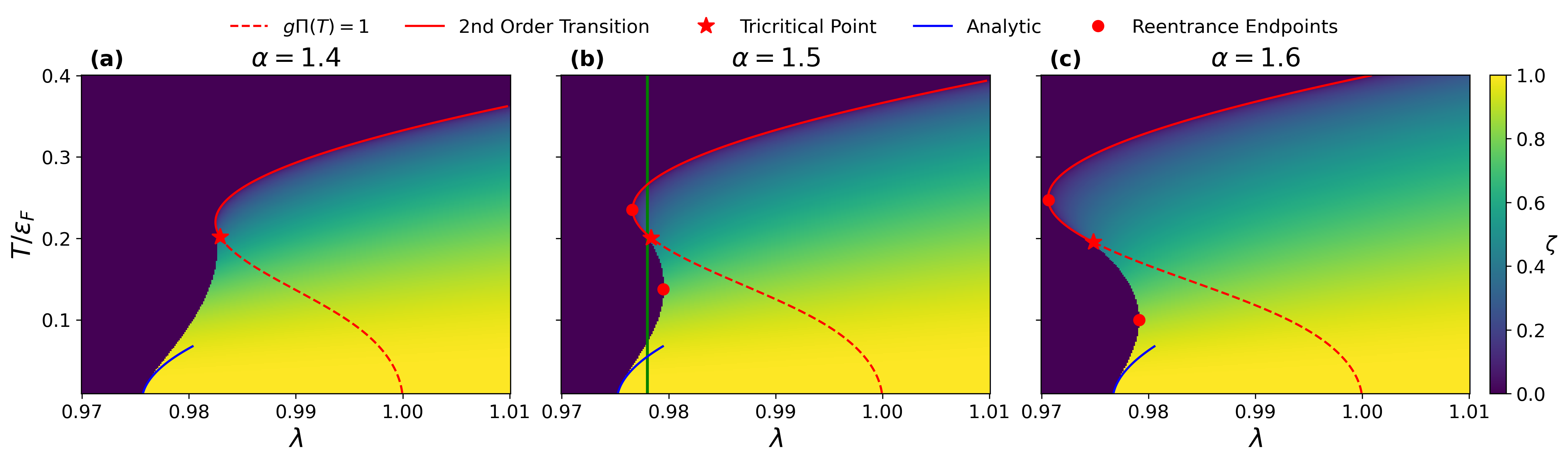}
		\caption{Phase diagrams for (a) $\alpha = 1.4$, (b) $\alpha = 1.5$, and (c) $\alpha = 1.6$. The solid red line indicates where the transition is second order, and the dashed red line indicates where susceptibility diverges. The blue line denotes the low temperature expansion of the phase boundary. Reentrant behavior is evident in (b) and (c), where the system transitions from the normal state to the ordered as temperature is increased.}
		\label{reentrant}
	\end{center}
\end{figure}

We plot the phase diagram for $\alpha = 1.4,1.5,$ and $1.6$ in Fig. \ref{reentrant}. At low temperatures, the $T^2$ expansion accurately describes the phase boundary and $\lambda_c$ increases with temperature. However, as temperature gets larger, the phase boundary bends away from the low temperature expansion. At sufficiently high temperatures, the phase transition becomes second order, and the phase boundary is set by the divergence of the  susceptibility. 
We mark the location of the tricritical point, i.e. the point where the first-order transition becomes a second-order transition by a star.

Notably, in these three figures, one can see that reentrant behavior emerges for $\alpha>1.4$, where the system transitions from the normal state to the ordered state at some finite temperature. For example, at $\alpha=1.5$, one can fix $\lambda =.978$ and increase temperature (see the green vertical line in Fig. \ref{reentrant} (b)).
 The system initially resides in the ordered state at low temperatures, transitions to the normal state at some intermediate temperature, reenters the ordered phase again at a higher temperature and eventually transition to the normal state at even higher $T$. A similar trend is observed for $\alpha =1.6$ in Fig. \ref{reentrant} (c).

\subsection{$2 \geq \alpha$}
We first consider $\alpha =2$ and then $\alpha \geq 2$. As before, we present the details of calculations in Appendix \ref{alpha=2} and list the results here. We find for the critical interaction strength,
\begin{align}
\lambda_c
(T)  &= 1 - 0.377 \left(\frac{T}{\e_F}\right)^{\frac{3}{2}}+ \mathcal{O}\left( T^{2} \right).
\end{align}
This  $\lambda_c (T) $ matches the divergence of the susceptibility at $T=0$. At a finite $T$, the susceptibility diverges at
\begin{align}
\lambda_c^{\chi} = 1 - \frac{\pi^2}{12} \left(\frac{T}{\e_F}\right)^2 > \lambda_c (T)
\end{align}
(see Eq. (\ref{susceptibility})). This indicates that there is a first order phase transition at sufficiently low temperatures.

\begin{figure}[h]
	\begin{center}
		\includegraphics[scale=.7]{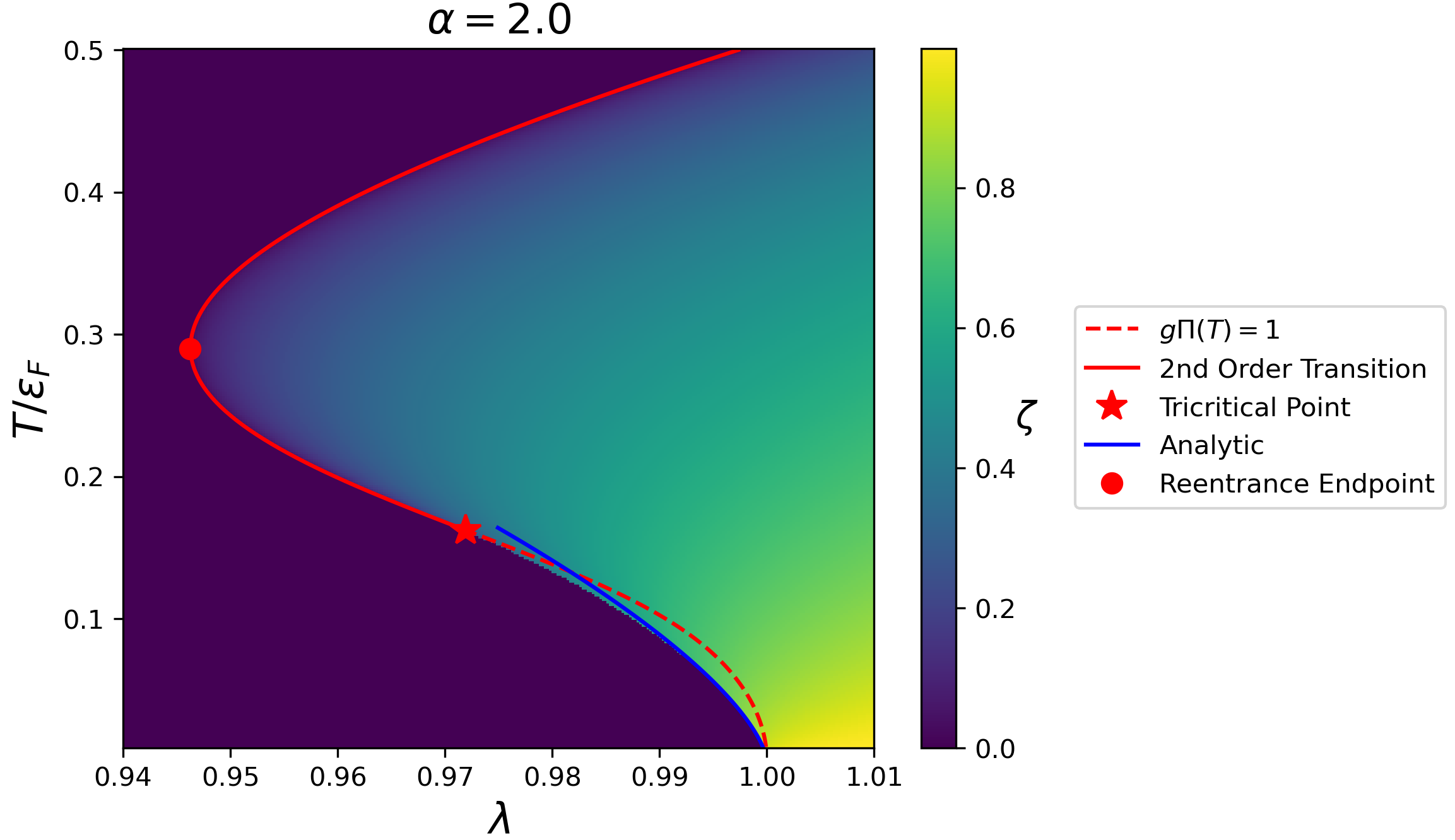}
		\caption{The phase diagram for $\alpha =2$, i.e. $\e_k \propto k^4$. At the lowest temperatures, the transition is first order, with $\lambda_c = 1 - 0.377 \left(T/\e_F\right)^{\frac{3}{2}}$, plotted in blue. As previously, we mark the second-order phase boundary with a solid red line, while the diverging susceptibility not associated with a phase transition is marked with a dashed line.}
		\label{quartic}
	\end{center}
\end{figure}

We compute the phase diagram numerically and show the results in Fig \ref{quartic}. We see that there is, indeed, a first-order phase transition up to $T \sim .15\e_F$ and a second-order transition at larger $T$.

\begin{figure}[h]
	\begin{center}
		\includegraphics[scale=.44]{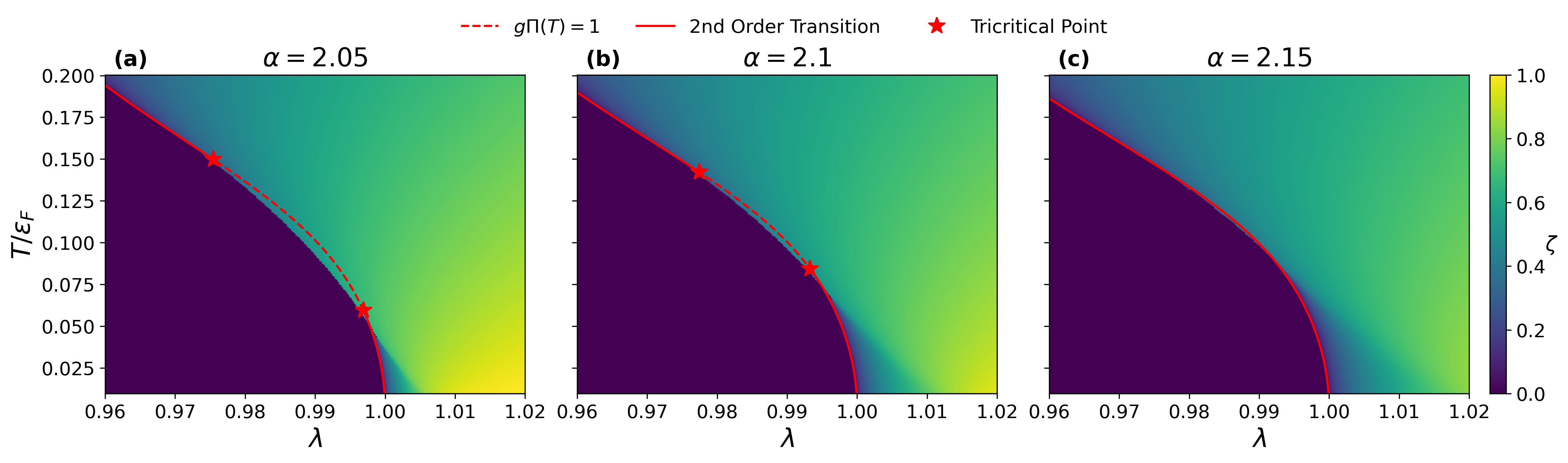}
		\caption{Phase diagram for (a) $\alpha=2.05$, (b) $\alpha=2.1$, and (c) $\alpha =2.15$. The range of temperatures over which the transition is first order decreases as $\alpha$ increases, disappearing at $\alpha =2.15$. However, the reentrant behavior is consistently seen at all $\alpha$.}
		\label{alpha>2}
	\end{center}
\end{figure}

For $\alpha>2$ the system at $T=0$ goes through a conventional second order phase transition. At the lowest temperatures, this is still the case. However, for $\alpha \lesssim 2.15 $, there still exists an intermediate range of temperatures in which the transition is first order. The lower bound of this range emerges from $T=0$ as $\alpha$ exceeds $2$, the upper boundary was finite already at $\alpha =2$. As $\alpha$ increases towards $2.15$, the two boundaries come close to each other and  merge at $\alpha =2.15$. We illustrate this behavior in Fig. \ref{alpha>2} where we plot phase diagrams for $\alpha = 2.05,2.1,$ and $2.15$.

\begin{figure}[h]
	\begin{center}
		\includegraphics[scale=.7]{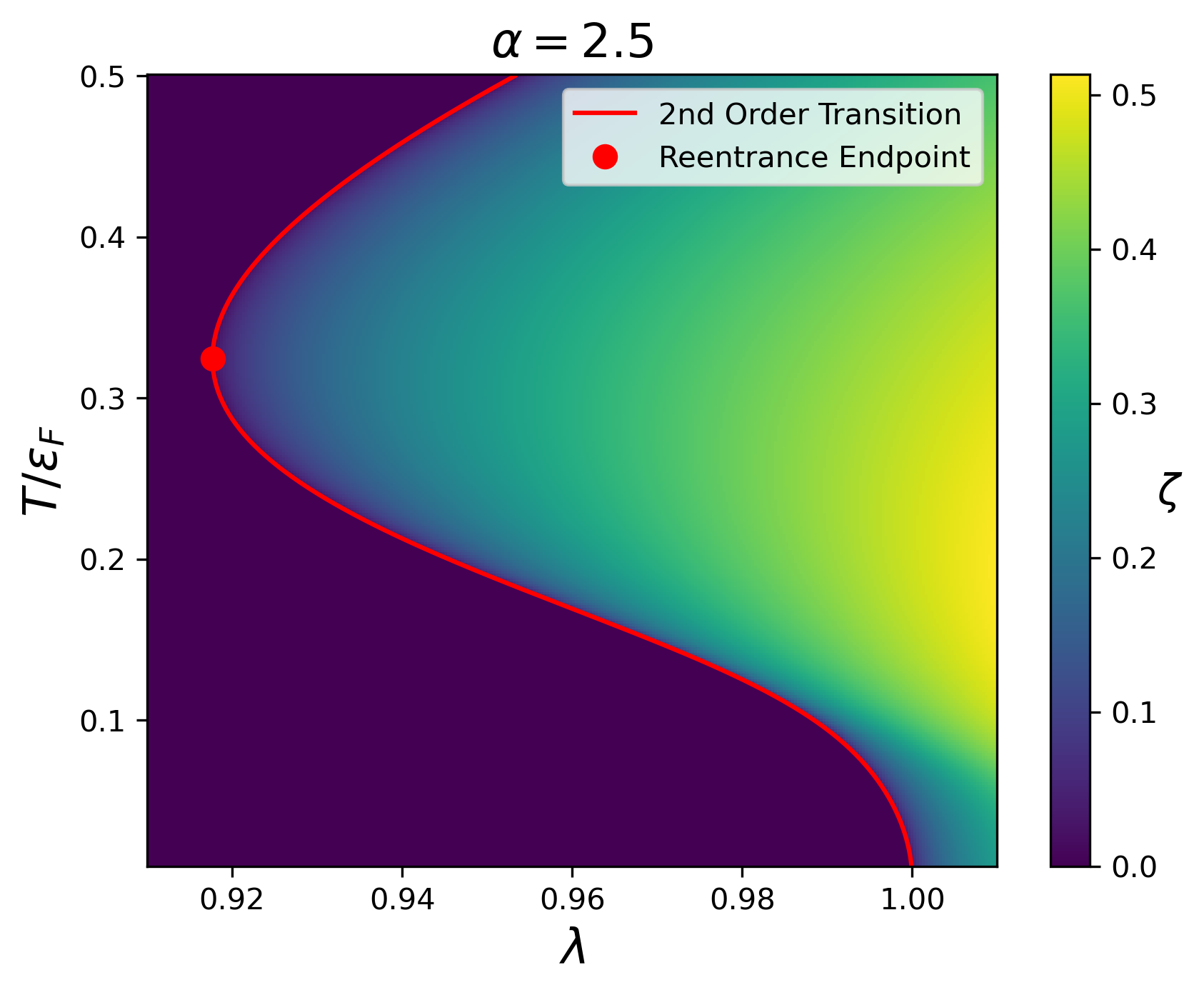}
		\caption{The phase diagram for $\alpha=2.5$, i.e. $\e_k \propto k^5$. The transition is second order throughout the phase diagram, and the reentrant behavior is seen at a wide range of interactions, occurring between $\lambda = 0.92$ and $\lambda = 1$.}
		\label{quintic}
	\end{center}
\end{figure}

For $\alpha \gtrsim 2.15$, the transition is second order throughout the phase diagram. However, the reentrant behavior remains intact and becomes more pronounced as $\alpha$ increases and the dispersion becomes more flat at small $k$. As an example, we present the phase diagram for $\alpha = 2.5$ in Fig. \ref{quintic}.

\begin{figure}[h]
	\begin{center}
		\includegraphics[scale=.8]{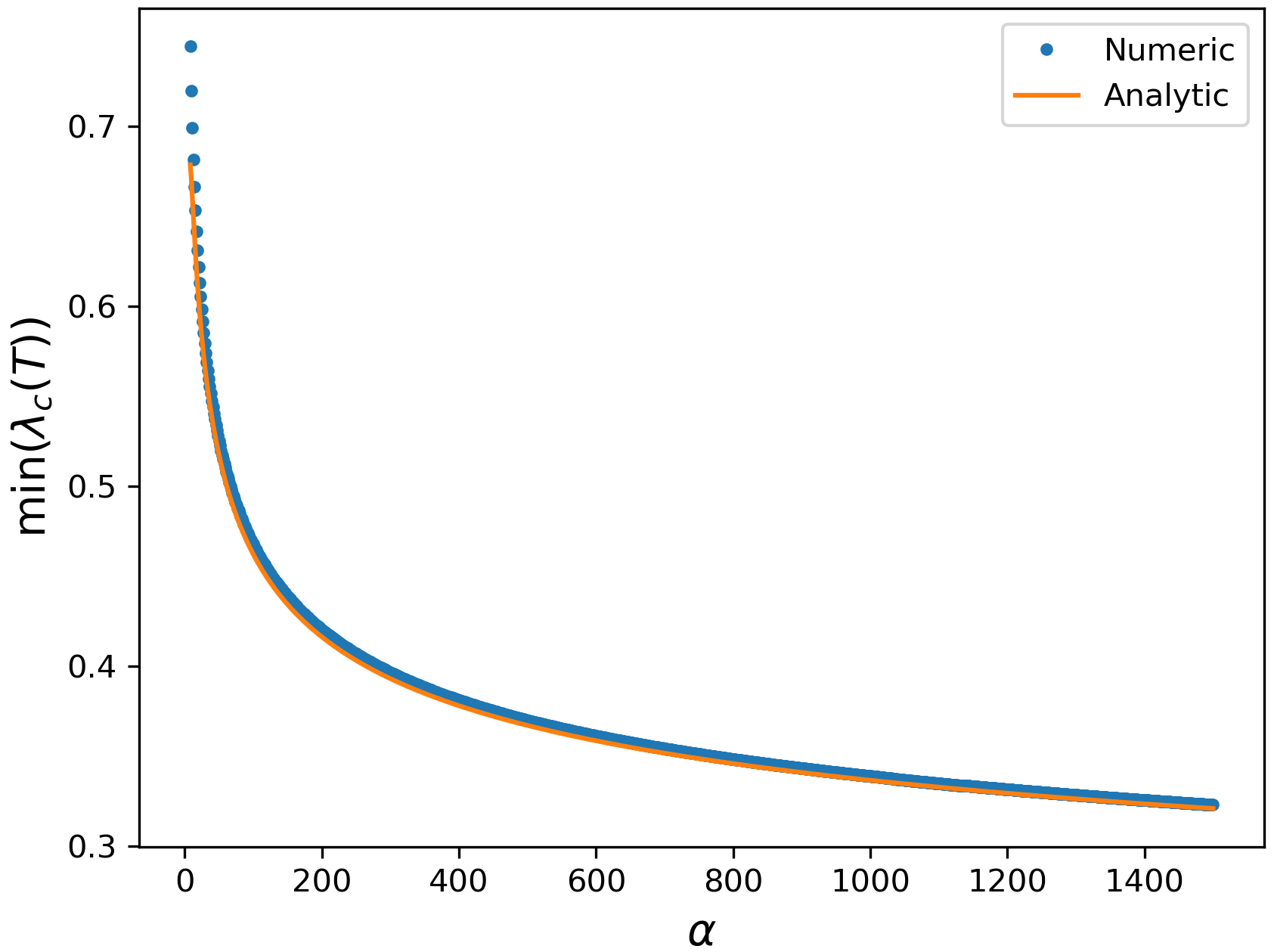}
		\caption{The minimum value of $\lambda_c(T)$ as a function of $\alpha$, evaluated both numerically and analytically. }
		\label{lambda_min}
	\end{center}
\end{figure}

We lastly discuss the case of $\alpha \rightarrow \infty$. As we increase $\alpha$, the range of $\lambda$ where the system exhibits reentrant behavior expands. To determine the minimum value of $\lambda_c(T)$, we search for the maximum value of $\Pi_{\alpha}(T)$, since $\lambda_c(T) = \frac{\nu_{F,\alpha}}{\Pi_{\alpha}(T)}$. We show in Appendix \ref{infinity} that the minimum $\lambda_c (T)$ does indeed go to zero at $\alpha \to \infty$ as
\begin{align}
\label{min_lambda} \text{min}\left(\lambda_{c}(T)\right) = \frac{e}{\log \alpha}\left(1-\frac{1}{\log \alpha}\right).
\end{align}
We plot the minimum value of $\lambda_c(T)$ as a function of $\alpha$ in Fig. \ref{lambda_min}. We plot both  Eq. (\ref{min_lambda}) and the result obtained by numerically evaluating the chemical potential and the polarization bubble. We find reasonable agreement between the two.

\section{Conclusions}
\label{conclusion}
In this paper, we have analyzed the Stoner transition in a 2D isotropic system with dispersion $\e_k \propto k^{2\alpha}$ at finite temperature. We demonstrated that the nature of the order parameter as well as the topology of the phase diagram depend strongly on the value of $\alpha$. For $\alpha<1$, the finite temperature behavior is the same as at $T=0$. That is, the transition is second order throughout the phase diagram. For $\alpha=1$, the parabolic dispersion, the phase transition is also second order at any finite temperatures. We recover the previous result at $T=0$ that the onset of the ordered phase coincides with the divergence of the susceptibility, despite the order parameter evolving discontinuously.

In the range $1<\alpha<2$, the transition is second order at higher temperatures and becomes first order at low temperatures. For $\alpha \gtrsim 1.4$, the phase boundary exhibits reentrant behavior, with the transitions from the normal state to the ordered state and then back to the normal state occurring as temperature is increased. The reentrant behavior becomes more prevalent as $\alpha$ is increased. For $\alpha \geq  2$, the transition is second order at low at and high temperatures and first order at intermediate temperatures.  The intermediate range shrinks as $\alpha$ increases, and vanishes for $\alpha \sim 2.15$. For larger $\alpha$, the transition is second order throughout the phase diagram. In the limit $\alpha \rightarrow \infty$, when the fermionic dispersion is essentially flat at small $k$, the phase diagram displays finger-type extension to the smallest $\lambda$.

We hope that our results, particularly the one about the reentrant behavior, will be relevant for non-twisted graphene multilayers.  In the simplest tight binding model, the low energy band structure takes the form $\e_k \sim k^n$, where $n$ is the number of graphene layers. The effective interaction $\lambda$ in these systems can be tuned via the density of states by modifying particle number. That said, further work would be needed to determine the full phase diagram for an experimentally relevant two valley system. In addition, including trigonal warping terms in the dispersion will most likely affect the polarization of the isospins, as has already been determined at zero temperature for bilayer graphene~\cite{mayrhofer2025}.

\section{Acknowledgments}
This work was supported by U.S. Department of Energy, Office of Science, Basic Energy Sciences, under Award No. DE-SC0014402. We thank E. Berg, A. Cherman and Z. Raines for useful discussions.

\appendix

\section{Polarization Bubble and Chemical Potential}

We present here the expansion of the chemical potential in the normal state as well as the polarization bubble at low temperatures, i.e. Eqs. (\ref{norm_chem}) and (\ref{susceptibility_t2}) in the main text. We start by considering the conservation particle number in the normal state,
\begin{align}
2\alpha \nu_{F,\alpha} \e_F =
- 2 \nu_{F,\alpha}\e_F \left(\frac{T}{\e_F}\right)^{\frac{1}{\alpha}} \Gamma\left(\frac{1}{\alpha}\right)\text{Li}_{\frac{1}{\alpha}} \left(-e^{\mu/T} \right).
\end{align}
Simplifying this expansion, and using the expansion for the polylogarithm in Eq. (\ref{polylog}), we obtain
\begin{align}
\alpha &= \left(\frac{T}{\e_F}\right)^{\frac{1}{\alpha}} \Gamma\left(\frac{1}{\alpha}\right) \left(\frac{1}{\Gamma\left(\frac{1}{\alpha}+1\right)} \left(\frac{\mu}{T} \right)^{\frac{1}{\alpha}} +\frac{\pi^2}{6}\frac{1}{\Gamma\left(\frac{1}{\alpha}-1\right)} \left(\frac{\mu}{T} \right)^{\frac{1}{\alpha}-2} \right) \\
&= \alpha \left(\frac{\mu}{\e_F}\right)^{\frac{1}{\alpha}} - \frac{\left(\alpha - 1\right) \pi^2}{6\alpha} \mu^{\frac{1}{\alpha}-2} \left(\frac{T}{\e_F}\right)^2.
\end{align}
Keeping only terms up to order $T^2$, we can then solve for $\mu$ to obtain the expression in Eq. (\ref{norm_chem}),
\begin{align}
\frac{\mu(T)}{\e_F} = 1+\frac{\pi^2}{6} \frac{\left(\alpha -1\right)}{\alpha}\left(\frac{T}{\e_F}\right)^2.
\end{align}
We can then insert this into polarization bubble,
\begin{align}
\Pi_{\alpha}(T) &= - \nu_{F,\alpha}\Gamma \left(\frac{1}{\alpha}\right) \left(\frac{T}{\e_F}\right)^{\frac{1}{\alpha} -1} \text{Li}_{\frac{1}{\alpha}-1}\left(-e^{\mu/T}\right) \\
&= \nu_{F,\alpha} \Gamma \left( \frac{1}{\alpha} \right) \left(\frac{T}{\e_F}\right)^{\frac{1}{\alpha}-1} \left( \frac{1}{\Gamma \left(\frac{1}{\alpha}\right)} \left(\frac{\mu}{T} \right)^{\frac{1}{\alpha}-1} +\frac{\pi^2}{6} \frac{1}{\Gamma \left(\frac{1}{\alpha}-2\right)} \left(\frac{\mu}{T} \right)^{\frac{1}{\alpha}-3}  \right).
\end{align}
Inserting our expression for $\mu(T)$ into this expression, we obtain the same expression as in Eq. (\ref{susceptibility_t2}),
\begin{align}
\Pi_{\alpha}(T) = \nu_{F,\alpha} \left(1 + \frac{\pi^2}{6} \frac{\left(\alpha -1\right)}{\alpha}\left(\frac{T}{\e_F}\right)^2 + \mathcal{O}\left(T^4\right) \right).
\end{align}

\section{Calculation of Order Parameter and Phase Boundaries}

\subsection{$\alpha = 1$}
\label{alpha=1}
We present here the calculation of the order parameter for $\alpha = 1$. We first consider $\lambda \sim \lambda_c(T)$, which will yield Eq. (\ref{small_quadratic_analytic}) in the main text. We note that for $\alpha=1$ the self-consistent equations are simplified since $\text{Li}_{1}(z) = - \log(1-z)$, and we can write
\begin{align}
\label{alpha1delta} \frac{\Delta}{\lambda} &= \frac{T}{2} \log \left( \frac{1+e^{\left(\bar \mu + \Delta\right)/T}}{1+e^{\left(\bar \mu - \Delta\right)/T}}\right),\\
\label{alpha1chem} \e_F &= \frac{T}{2}\log \left( \left(1+e^{\left(\bar \mu + \Delta\right)/T}\right)\left(1+e^{\left(\bar \mu - \Delta\right)/T}\right)\right).
\end{align}
To begin, we expand Eq. (\ref{alpha1chem}) to order $\Delta^2$,
\begin{align}
\e_F = T\log \left(1+e^{\frac{\bar \mu}{T}} \right) + \frac{e^{\frac{\bar \mu}{T}} \Delta^2}{2T\left(1+e^{\frac{\bar \mu}{T}}\right)}.
\end{align}
Then, we can solve for $\bar \mu$ up to order $\Delta^2$,
\begin{align}
\bar \mu= T \log \left(e^{\frac{\e_F}{T}} - 1 \right) - \frac{e^{-\frac{\e_F}{T}}}{2T} \Delta^2.
\end{align}
We can see here that the corrections to the chemical potential due to a finite $\Delta$ are suppressed by a factor $e^{-\frac{\e_F}{T}}$. We can then safely set $\Delta =0$ in $\bar \mu$ when determining $\Delta$ at low temperatures in the vicinity of the quantum critical point. Expanding Eq. (\ref{alpha1delta}) to order $\Delta^3$, we find
\begin{align}
\frac{\Delta}{\lambda} =  \frac{\Delta e^{\frac{\bar \mu}{T}}}{1+e^{\frac{\bar \mu}{T}}} \left( 1-  \frac{e^{\frac{\bar \mu}{T}}-1}{6\left(e^{\frac{\bar \mu}{T}}+1\right)^2}\frac{\Delta^2}{T^2} \right).
\end{align}
Inserting the expression for $\bar \mu$ we obtained above and solving for $\Delta^2$, we find
\begin{align}
\Delta^2 = \frac{6T^2 e^{\frac{\e_F}{T}}}{\left( 1 - 2e^{-\frac{\e_F}{T}}\right)} \frac{\lambda - \lambda_c(T)}{\lambda} \sim
6T^2 e^{\frac{\e_F}{T}}\left( \lambda - \lambda_c(T)\right),
\end{align}
where $\lambda_c(T) = (1-e^{-\frac{\e_F}{T}})^{-1}$, and we have assumed $T\ll\e_F$, $\lambda \sim \lambda_c(T)$, and kept only terms leading order $e^{\frac{\e_F}{T}}$.

We now present the calculation for $\Delta$ close to its saturated value, i.e. the expression in Eq. (\ref{quadratic_analytic}). We can add and subtract Eq. (\ref{alpha1delta}) and (\ref{alpha1chem}) to obtain two simpler equations for $\Delta$ and $\bar \mu$,
\begin{align}
\e_F + \frac{\Delta}{\lambda} &= T \log \left( 1+ e^{(\bar \mu + \Delta)/T}\right), \\
\e_F -  \frac{\Delta}{\lambda} &= T \log \left( 1+ e^{(\bar \mu - \Delta)/T}\right).
\end{align}
Solving the first equation for $\bar \mu$, we find
\begin{align}
\bar \mu = T \log \left(e^{\left(\e_F +  \frac{\Delta}{\lambda}\right)/T} - 1\right) -\Delta.
\end{align}
Inserting this into the second equation, we find
\begin{align}
e^{(\e_F-\frac{\Delta}{\lambda})/T} - e^{\left(\e_F+\Delta \left(\frac{1}{\lambda} -2 \right) \right)/T} + e^{-\frac{2\Delta}{T}} = 1.
\end{align}
Here, we will make the ansatz that $\Delta/\lambda = \e_F - \delta$, and we will assume that we are in some vicinity of the critical point $\epsilon = \lambda -1 > 0$. The above equation is then written as
\begin{align}
1 &= e^{\delta/T} - e^{\left(\delta - 2\e_F \epsilon + 2 \delta \epsilon \right)/T} + e^{-2 \left(\e_F-\delta\right)/T}.
\end{align}

Assuming that $\delta \ll \e_F$, we can then keep leading order terms in $\delta$ to obtain
\begin{align}
\delta = - T \log \left( 1- e^{-\frac{2 \e_F \epsilon}{T}} - e^{-\frac{2\e_F}{T}}\right)
\implies \frac{\Delta}{\lambda}  = \e_F + T \log \left( 1- e^{-\frac{2\e_F\left(\lambda -1 \right)}{T}} -e^{-\frac{2\e_F}{T}} \right).
\end{align}
For $\lambda < 2$ and $T\ll\e_F$, we can see $e^{-\frac{2\e_F\left(\lambda -1 \right)}{T}}\gg e^{-\frac{2\e_F}{T}}$, so we can neglect the $e^{-\frac{2\e_F}{T}}$ term in the logarithm when in the vicinity of the quantum critical point. Therefore, we arrive at the expression in Eq. (\ref{quadratic_analytic}) in the main text.

\subsection{$1<\alpha<2$}
\label{alpha>1}
We next present the calculations of the results for $1<\alpha<2$. To start, we add and subtract Eq. (\ref{eq1}) and (\ref{particle_number}) to obtain two new equations,
\begin{align}
\label{delta1} &\alpha + \frac{\Delta}{\lambda \e_F}= -\left(\frac{T}{\e_F}\right)^{\frac{1}{\alpha}} \Gamma \left(\frac{1}{\alpha}\right) \text{Li}_{\frac{1}{\alpha}} \left( -e^{\left(\bar \mu + \Delta \right)/T}\right),\\
\label{delta2} &\alpha -\frac{\Delta}{\lambda \e_F}  = -\left(\frac{T}{\e_F}\right)^{\frac{1}{\alpha}}  \Gamma \left(\frac{1}{\alpha}\right) \text{Li}_{\frac{1}{\alpha}} \left( -e^{\left(\bar \mu - \Delta \right)/T}\right).
\end{align}
To do this, we will first solve for the chemical potential of the system by examining Eq. (\ref{delta1}). We can expand the polylogarithm in this equation by using the series expansion in Eq. (\ref{polylog}). Solving for the chemical potential up to order $T^2$, we find
\begin{align}
\label{chem} \frac{\bar \mu}{\e_F} = \left(1 + \frac{\Delta}{\alpha \lambda \e_F}\right)^{\alpha} - \frac{\Delta}{\e_F} +\frac{\pi^2}{6} \frac{\alpha-1}{\alpha} \left(1 + \frac{\Delta}{\alpha \lambda \e_F}\right)^{-\alpha}\left( \frac{T}{\e_F}\right)^2.
\end{align}
We can then solve Eq. (\ref{delta2}). At $T=0$, we will assume and later verify that $\Delta>\bar \mu$, so $e^{(\bar \mu - \Delta)/T}$ is exponentially small at low temperatures. For any $s>0$, $\text{Li}_{s} (z) \sim z$ for $z \ll 1$. The corrections to the polylogarithm are then exponentially small. We neglect these exponential corrections, as these will always be smaller than a power law correction at low temperatures. Therefore, we can safely set the right hand side of Eq. (\ref{delta2}) to zero. We can then easily see that $\Delta = \alpha \lambda \e_F + \mathcal{O}\left( e^{\frac{\bar \mu - \alpha \lambda}{T}}\right)$ and that the chemical potential can be written as
\begin{align}
\frac{\bar \mu(T)}{\e_F} = 2^{\alpha} - \alpha \lambda +  \frac{\left(\alpha-1\right) \pi^2}{3 \alpha}\frac{1}{2^{1+\alpha}} \left(\frac{T}{\e_F}\right)^2.
\end{align}
At $T=0$, $\bar \mu - \Delta = \e_F \left(2^{\alpha} - 2\alpha \lambda \right)$. In the ordered state, for $1<\alpha<2$, we can see that $\bar \mu - \Delta <0$, so our initial assumption is justified. We can then insert these expressions for $\bar \mu$ and $\Delta$ into the free energy Eq. (\ref{freeenergy}). To determine where the onset of the ordered state occurs, we must subtract off the normal state free energy, i.e. Eq. (\ref{freeenergy}) with $\Delta = 0$. Doing so and expanding to order $T^2$, we obtain for the free energy
\begin{align}
\frac{\delta F}{N} = \frac{F_{ord}-F_N}{N} = \frac{\alpha \e_F}{2} \left( \frac{2\left(2^{\alpha}-1\right)}{\alpha\left(1+\alpha\right)} - \lambda\right) + \frac{\pi^2}{6\alpha} \left(1-2^{-\alpha}\right)\frac{T^2}{\e_F},
\end{align}
where $F_N$ is the free energy in the normal state. Setting this equal to zero, we can solve for the critical $\lambda$ to obtain
\begin{align}
\lambda_c = \frac{2\left(2^\alpha-1\right)}{\alpha \left(1+\alpha \right)} + \frac{\left(1-2^{-\alpha}\right)}{3\alpha^2} \pi^2 \left(\frac{T}{\e_F}\right)^2.
\end{align}

\subsection{$\alpha = 2$}
\label{alpha=2}
We present the calculation of the critical interaction strength for $\alpha=2$. To start, we make the ansatz
\begin{align}
\frac{\Delta}{\lambda} = 2\e_F - A \sqrt{\e_F T}.
\end{align}
In addition, we assume that $\lambda = 1$, since this is the location of the phase transition at $T=0$. The critical $\lambda$ will vary with temperature, however we will show that the corrections to the order parameter from the temperature dependence of $\lambda $ are subleading. We also  note that, since $\bar \mu + \Delta >0$ at low temperatures, the expansion in Eq. (\ref{chem}) is still valid. Then, using this ansatz for $\Delta$ and Eq. (\ref{chem}), we write
\begin{align}
&\bar \mu - \Delta = \frac{A^2}{4}T.
\end{align}
Inserting this into Eq. (\ref{delta2}) to obtain an equation for $A$,
\begin{align}
A = -\sqrt{\pi} \text{Li}_{\frac{1}{2}}\left(-e^{\frac{A^2}{4}}\right),
\end{align}
where we have used $\Gamma(1/2) = \sqrt{\pi}$. Numerically solving this equation, we obtain $A \sim 1.4525$. To determine the critical interaction, we substitute these results into the free energy for the system in Eq. (\ref{freeenergy}), and expand to leading order in temperature to obtain
\begin{align}
\frac{\delta F}{N \e_F} = 1-\lambda -\left(1-\lambda\right)  \frac{A^2 T}{4\e_F} + \frac{1}{4}\left(\frac{A^3}{6} + \sqrt{\pi} \text{Li}_{\frac{3}{2}} \left(-e^{\frac{A^2}{4}} \right) \right) \left(\frac{T}{\e_F}\right)^{\frac{3}{2}} + \mathcal{O}\left(T^2\right).
\end{align}
We write $\lambda = 1 - \epsilon$ and solve for $\lambda$ such that the free energy in the ordered and normal states are equal. Setting $\delta F=0$ and expanding in small $\epsilon$, we determine that, to leading order in temperature
\begin{align}
\epsilon &= -\frac{1}{4} \left(\frac{A^3}{6} + \sqrt{\pi} \text{Li}_{\frac{3}{2}} \left( - e^{\frac{A^2}{4}}\right) \right) \left(\frac{T}{\e_F}\right)^{\frac{3}{2}} \sim 0.377 \left(\frac{T}{\e_F}\right)^{\frac{3}{2}}\\
\implies \lambda_c &= 1 - 0.377 \left(\frac{T}{\e_F}\right)^{\frac{3}{2}}.
\end{align}
We note that in the initial ansatz for $\Delta$, we neglected the temperature dependence in $\lambda_c$. We can see that this assumption is justified, as the leading order correction in $\bar \mu -\Delta$ is linear in $T$, whereas corrections arising from the temperature dependance of $\lambda_c$ are proportional to $T^{\frac{3}{2}}$.

\subsection{$\alpha \rightarrow \infty$}
\label{infinity}
We analyze the critical coupling in the limit $\alpha \rightarrow \infty$, and show how to obtain the minimum value of $\lambda_c(T)$ as written in Eq. (\ref{min_lambda}). Since the transition is second order, it is sufficient to calculate the maximum value of polarization bubble as a function of $T$. However, to calculate $\Pi_{\alpha}(T)$, we first must determine the chemical potential of the system at large $\alpha$. To do so, we begin by considering equation for conservation of number of particles in the normal state,
\begin{align}
2 \alpha  \nu_{F,\alpha} \e_F &= 2 \nu_{F,\alpha} \e_F^{1-\frac{1}{\alpha}} \int \limits_{0}^{\infty} d \e_k \e_k^{\frac{1}{\alpha} -1} n_F\left(\e_k - \mu\right), \\
\label{a41} \implies \alpha &= \left(\frac{T}{\e_F}\right)^{\frac{1}{\alpha}} I,
\quad I =  \int \limits_{0}^{\infty} dx \frac{x^{\frac{1}{\alpha}-1}}{1+e^{x-\frac{\mu}{T}}},
\end{align}
where we have written $x = \e_k/T$. We evaluate the integral in $I$ by parts to obtain
\begin{align}
I = \frac{\alpha x^{\frac{1}{\alpha}}}{1+e^{x-\frac{\mu}{T}}} \Big \rvert_{0}^{\infty} + \int \limits_{0}^{\infty} \frac{\alpha x^{\frac{1}{\alpha}}}{4 \cosh^2\left(\frac{\left(x-\mu/T\right)}{2} \right)}.
\end{align}
We note that, for any $\alpha>0$, the first term will evaluate to zero. To evaluate the second term, we can then expand in large $\alpha$,
\begin{align}
I &= \int \limits_{0}^{\infty} \frac{\alpha + \log(x)}{4 \cosh^2\left(\frac{\left(x-\mu/T\right)}{2} \right)} + \mathcal{O}\left(\alpha^{-1}\right) \\
&= \frac{\alpha}{1+e^{-\frac{\mu}{T}}} + \int \limits_{0}^{\infty} \frac{\log(x)}{4 \cosh^2\left(\frac{\left(x-\mu/T\right)}{2} \right)} + \mathcal{O}\left(\alpha^{-1}\right).
\end{align}
We note that $\cosh(x)^2$ is peaked around $x=0$, so to logarithmic accuracy, we can write $I$ as
\begin{align}
I = \frac{\alpha}{1+e^{-\frac{\mu}{T}}} + \frac{\log\left(\frac{\mu}{T}\right)}{1+e^{-\frac{\mu}{T}}}.
\end{align}
Substituting this result into Eq. (\ref{a41}) and expanding $T^{\frac{1}{\alpha}}$ to leading order in $\alpha$, we obtain
\begin{align}
\alpha = \left(1+\frac{\log \frac{T}{\e_F}}{\alpha}\right) \left(\alpha + \log\left(\frac{\mu}{T}\right)\right) \frac{1}{1+e^{-\frac{\mu}{T}}}.
\end{align}
Neglecting the term proportion to $\alpha^{-1}$, we find for the chemical potential
\begin{align}
\label{mu} \frac{\mu}{T} =  \log \alpha - \log \left(\log \left( \frac{\mu}{\e_F} \right)\right).
\end{align}
We can then iteratively solve for $\mu$ to leading order. Doing so, we obtain
\begin{align}
\frac{\mu}{T} = \log \alpha - \log \left( \log \left( \frac{T}{\e_F} \log \alpha \right) \right).
\end{align}
Now, we can directly calculate the polarization bubble,
\begin{align}
\Pi_{\alpha}(T) = - \nu_{F,\alpha} \e_F^{1-\frac{1}{\alpha}}\int \limits_{0}^{\infty} d\e_k \e_k^{\frac{1}{\alpha}-1} n_F'(\e_k-\mu).
\end{align}
We proceed in a manner similar to what we did to calculate the chemical potential. We rescale $\e_k$ by $T$, integrate by parts, and then expand $x^{\frac{1}{\alpha}}$ to leading order in $\alpha$. Doing so, we obtain
\begin{align}
\frac{\Pi_{\alpha}(T)}{\nu_{F,\alpha}} &= \frac{\e_F}{4T} \left(1 + \frac{\log \frac{T}{\e_F}}{\alpha}\right) \int \limits_{0}^{\infty} dx \frac{\alpha + \log(x)}{\cosh^2\left( \frac{x - \mu/T}{2}\right) \coth \left( \frac{x - \mu/T}{2}\right) } \\
 &=\frac{\e_F}{4T} \left(1 + \frac{\log \frac{T}{\e_F}}{\alpha}\right) \left(\alpha \sech^2 \left(\frac{\mu}{2T}\right) + \int \limits_{-\mu/T}^{\infty} dx \frac{ \log(x+\frac{\mu}{T})}{\cosh^2\left( \frac{x}{2}\right) \coth \left( \frac{x}{2}\right) }\right).
\end{align}
We can expand log in powers of $x$, so $\log \left( x + \frac{\mu}{T} \right) \sim \log \left( \frac{\mu}{T} \right) + \frac{Tx}{\mu}$. However, we can quickly see that the term with $\log \left( \frac{\mu}{T} \right)$ is exponentially suppressed by a factor $e^{-\frac{\mu}{T}}\sim \alpha^{-1}$, since the integrand is odd over $x$. We neglect this term, and instead include only the next order term with the lower bound set to $-\infty$, since $\mu/T \sim \log \alpha$. We can therefore write the above expression for $\Pi_{\alpha}(T)$ as
\begin{align}
\frac{\Pi_{\alpha}(T)}{\nu_{F,\alpha}} &=\frac{\e_F}{4T} \left(1 + \frac{\log \frac{T}{\e_F}}{\alpha}\right) \left(\alpha \sech^2 \left(\frac{\mu}{2T}\right) + \frac{T}{\mu} \int \limits_{-\infty}^{\infty} dx \frac{x}{\cosh^2\left( \frac{x}{2}\right) \coth \left( \frac{x}{2}\right) }\right)\\
&= \frac{\e_F}{4T} \left(1 + \frac{\log \frac{T}{\e_F}}{\alpha}\right)  \left(\alpha \sech^2 \left(\frac{\mu}{2T}\right) + \frac{4T}{\mu} \right).
\end{align}
Now, we can insert the expression we had obtained for $\mu$ into this expression to obtain, to subleading order in $\alpha$
\begin{align}
\frac{\Pi_{\alpha}(T)}{\nu_{F,\alpha}} &= \frac{\e_F}{T} \left(\log \left(\frac{T}{\e_F}\log \alpha \right) + \frac{1}{\log \alpha}\right) + \mathcal{O}\left(\alpha^{-1}\right).
\end{align}
We can locate the temperature at which the polarization bubble is at a maximum by taking $\frac{\partial \Pi_{\alpha}(T)}{\partial T} = 0$. Doing this, we find,
\begin{align}
\frac{T_{m}}{\e_F} = \frac{e}{\log \alpha} \left( 1 - \frac{1}{\log \alpha} \right).
\end{align}

Reinserting this expression for the temperature into the polarization bubble, we obtain for the maximum value of the polarization bubble,
\begin{align}
\frac{\Pi_{\alpha}(T_{m})}{\nu_{F,\alpha}} = \frac{\log \alpha+1}{e}.
\end{align}
We can then write $\text{min}\left(\lambda_c(T)\right)$ to order $1/(\log\alpha)^2$ as
\begin{align}
\text{min}\left(\lambda_c(T)\right) = \frac{\nu_{F,\alpha}}{\Pi_{\alpha}(T_{m})} = \frac{e}{\log \alpha} \left( 1 - \frac{1}{\log \alpha} \right).
\end{align}
We can see that $\text{min}\left(\lambda_c(T)\right)$ decreases logarithmically as $\alpha$ increases, as presented in the main text.

\bibliography{stoner_bib}

\begin{thebibliography}{55}%
\makeatletter
\providecommand \@ifxundefined [1]{%
 \@ifx{#1\undefined}
}%
\providecommand \@ifnum [1]{%
 \ifnum #1\expandafter \@firstoftwo
 \else \expandafter \@secondoftwo
 \fi
}%
\providecommand \@ifx [1]{%
 \ifx #1\expandafter \@firstoftwo
 \else \expandafter \@secondoftwo
 \fi
}%
\providecommand \natexlab [1]{#1}%
\providecommand \enquote  [1]{``#1''}%
\providecommand \bibnamefont  [1]{#1}%
\providecommand \bibfnamefont [1]{#1}%
\providecommand \citenamefont [1]{#1}%
\providecommand \href@noop [0]{\@secondoftwo}%
\providecommand \href [0]{\begingroup \@sanitize@url \@href}%
\providecommand \@href[1]{\@@startlink{#1}\@@href}%
\providecommand \@@href[1]{\endgroup#1\@@endlink}%
\providecommand \@sanitize@url [0]{\catcode `\\12\catcode `\$12\catcode
  `\&12\catcode `\#12\catcode `\^12\catcode `\_12\catcode `\%12\relax}%
\providecommand \@@startlink[1]{}%
\providecommand \@@endlink[0]{}%
\providecommand \url  [0]{\begingroup\@sanitize@url \@url }%
\providecommand \@url [1]{\endgroup\@href {#1}{\urlprefix }}%
\providecommand \urlprefix  [0]{URL }%
\providecommand \Eprint [0]{\href }%
\providecommand \doibase [0]{https://doi.org/}%
\providecommand \selectlanguage [0]{\@gobble}%
\providecommand \bibinfo  [0]{\@secondoftwo}%
\providecommand \bibfield  [0]{\@secondoftwo}%
\providecommand \translation [1]{[#1]}%
\providecommand \BibitemOpen [0]{}%
\providecommand \bibitemStop [0]{}%
\providecommand \bibitemNoStop [0]{.\EOS\space}%
\providecommand \EOS [0]{\spacefactor3000\relax}%
\providecommand \BibitemShut  [1]{\csname bibitem#1\endcsname}%
\let\auto@bib@innerbib\@empty
\bibitem [{\citenamefont {Novoselov}\ \emph {et~al.}(2004)\citenamefont
  {Novoselov}, \citenamefont {Geim}, \citenamefont {Morozov}, \citenamefont
  {Jiang}, \citenamefont {Zhang}, \citenamefont {Dubonos}, \citenamefont
  {Grigorieva},\ and\ \citenamefont {Firsov}}]{novoselov2004}%
  \BibitemOpen
  \bibfield  {author} {\bibinfo {author} {\bibfnamefont {K.~S.}\ \bibnamefont
  {Novoselov}}, \bibinfo {author} {\bibfnamefont {A.~K.}\ \bibnamefont {Geim}},
  \bibinfo {author} {\bibfnamefont {S.~V.}\ \bibnamefont {Morozov}}, \bibinfo
  {author} {\bibfnamefont {D.}~\bibnamefont {Jiang}}, \bibinfo {author}
  {\bibfnamefont {Y.}~\bibnamefont {Zhang}}, \bibinfo {author} {\bibfnamefont
  {S.~V.}\ \bibnamefont {Dubonos}}, \bibinfo {author} {\bibfnamefont {I.~V.}\
  \bibnamefont {Grigorieva}},\ and\ \bibinfo {author} {\bibfnamefont {A.~A.}\
  \bibnamefont {Firsov}},\ }\bibfield  {title} {\bibinfo {title} {Electric
  field effect in atomically thin carbon films},\ }\href
  {https://doi.org/10.1126/science.1102896} {\bibfield  {journal} {\bibinfo
  {journal} {Science}\ }\textbf {\bibinfo {volume} {306}},\ \bibinfo {pages}
  {666} (\bibinfo {year} {2004})},\ \Eprint
  {https://arxiv.org/abs/https://www.science.org/doi/pdf/10.1126/science.1102896}
  {https://www.science.org/doi/pdf/10.1126/science.1102896} \BibitemShut
  {NoStop}%
\bibitem [{\citenamefont {Novoselov}\ \emph {et~al.}(2005)\citenamefont
  {Novoselov}, \citenamefont {Jiang}, \citenamefont {Schedin}, \citenamefont
  {Booth}, \citenamefont {Khotkevich}, \citenamefont {Morozov},\ and\
  \citenamefont {Geim}}]{novoselov2005}%
  \BibitemOpen
  \bibfield  {author} {\bibinfo {author} {\bibfnamefont {K.~S.}\ \bibnamefont
  {Novoselov}}, \bibinfo {author} {\bibfnamefont {D.}~\bibnamefont {Jiang}},
  \bibinfo {author} {\bibfnamefont {F.}~\bibnamefont {Schedin}}, \bibinfo
  {author} {\bibfnamefont {T.}~\bibnamefont {Booth}}, \bibinfo {author}
  {\bibfnamefont {V.}~\bibnamefont {Khotkevich}}, \bibinfo {author}
  {\bibfnamefont {S.}~\bibnamefont {Morozov}},\ and\ \bibinfo {author}
  {\bibfnamefont {A.~K.}\ \bibnamefont {Geim}},\ }\bibfield  {title} {\bibinfo
  {title} {Two-dimensional atomic crystals},\ }\href@noop {} {\bibfield
  {journal} {\bibinfo  {journal} {Proceedings of the National Academy of
  Sciences}\ }\textbf {\bibinfo {volume} {102}},\ \bibinfo {pages} {10451}
  (\bibinfo {year} {2005})}\BibitemShut {NoStop}%
\bibitem [{\citenamefont {Dean}\ \emph {et~al.}(2010)\citenamefont {Dean},
  \citenamefont {Young}, \citenamefont {Meric}, \citenamefont {Lee},
  \citenamefont {Wang}, \citenamefont {Sorgenfrei}, \citenamefont {Watanabe},
  \citenamefont {Taniguchi}, \citenamefont {Kim}, \citenamefont {Shepard} \emph
  {et~al.}}]{dean2010}%
  \BibitemOpen
  \bibfield  {author} {\bibinfo {author} {\bibfnamefont {C.~R.}\ \bibnamefont
  {Dean}}, \bibinfo {author} {\bibfnamefont {A.~F.}\ \bibnamefont {Young}},
  \bibinfo {author} {\bibfnamefont {I.}~\bibnamefont {Meric}}, \bibinfo
  {author} {\bibfnamefont {C.}~\bibnamefont {Lee}}, \bibinfo {author}
  {\bibfnamefont {L.}~\bibnamefont {Wang}}, \bibinfo {author} {\bibfnamefont
  {S.}~\bibnamefont {Sorgenfrei}}, \bibinfo {author} {\bibfnamefont
  {K.}~\bibnamefont {Watanabe}}, \bibinfo {author} {\bibfnamefont
  {T.}~\bibnamefont {Taniguchi}}, \bibinfo {author} {\bibfnamefont
  {P.}~\bibnamefont {Kim}}, \bibinfo {author} {\bibfnamefont {K.~L.}\
  \bibnamefont {Shepard}}, \emph {et~al.},\ }\bibfield  {title} {\bibinfo
  {title} {Boron nitride substrates for high-quality graphene electronics},\
  }\href@noop {} {\bibfield  {journal} {\bibinfo  {journal} {Nature
  nanotechnology}\ }\textbf {\bibinfo {volume} {5}},\ \bibinfo {pages} {722}
  (\bibinfo {year} {2010})}\BibitemShut {NoStop}%
\bibitem [{\citenamefont {Radisavljevic}\ \emph {et~al.}(2011)\citenamefont
  {Radisavljevic}, \citenamefont {Radenovic}, \citenamefont {Brivio},
  \citenamefont {Giacometti},\ and\ \citenamefont {Kis}}]{radisavljevic2011}%
  \BibitemOpen
  \bibfield  {author} {\bibinfo {author} {\bibfnamefont {B.}~\bibnamefont
  {Radisavljevic}}, \bibinfo {author} {\bibfnamefont {A.}~\bibnamefont
  {Radenovic}}, \bibinfo {author} {\bibfnamefont {J.}~\bibnamefont {Brivio}},
  \bibinfo {author} {\bibfnamefont {V.}~\bibnamefont {Giacometti}},\ and\
  \bibinfo {author} {\bibfnamefont {A.}~\bibnamefont {Kis}},\ }\bibfield
  {title} {\bibinfo {title} {Single-layer mos2 transistors},\ }\href@noop {}
  {\bibfield  {journal} {\bibinfo  {journal} {Nature nanotechnology}\ }\textbf
  {\bibinfo {volume} {6}},\ \bibinfo {pages} {147} (\bibinfo {year}
  {2011})}\BibitemShut {NoStop}%
\bibitem [{\citenamefont {Geim}\ and\ \citenamefont
  {Grigorieva}(2013)}]{geim2013}%
  \BibitemOpen
  \bibfield  {author} {\bibinfo {author} {\bibfnamefont {A.~K.}\ \bibnamefont
  {Geim}}\ and\ \bibinfo {author} {\bibfnamefont {I.~V.}\ \bibnamefont
  {Grigorieva}},\ }\bibfield  {title} {\bibinfo {title} {Van der waals
  heterostructures},\ }\href {https://doi.org/10.1038/nature12385} {\bibfield
  {journal} {\bibinfo  {journal} {Nature}\ }\textbf {\bibinfo {volume} {499}},\
  \bibinfo {pages} {419} (\bibinfo {year} {2013})}\BibitemShut {NoStop}%
\bibitem [{\citenamefont {Cao}\ \emph {et~al.}(2018{\natexlab{a}})\citenamefont
  {Cao}, \citenamefont {Fatemi}, \citenamefont {Fang}, \citenamefont
  {Watanabe}, \citenamefont {Taniguchi}, \citenamefont {Kaxiras},\ and\
  \citenamefont {Jarillo-Herrero}}]{cao2018}%
  \BibitemOpen
  \bibfield  {author} {\bibinfo {author} {\bibfnamefont {Y.}~\bibnamefont
  {Cao}}, \bibinfo {author} {\bibfnamefont {V.}~\bibnamefont {Fatemi}},
  \bibinfo {author} {\bibfnamefont {S.}~\bibnamefont {Fang}}, \bibinfo {author}
  {\bibfnamefont {K.}~\bibnamefont {Watanabe}}, \bibinfo {author}
  {\bibfnamefont {T.}~\bibnamefont {Taniguchi}}, \bibinfo {author}
  {\bibfnamefont {E.}~\bibnamefont {Kaxiras}},\ and\ \bibinfo {author}
  {\bibfnamefont {P.}~\bibnamefont {Jarillo-Herrero}},\ }\bibfield  {title}
  {\bibinfo {title} {Unconventional superconductivity in magic-angle graphene
  superlattices},\ }\href {https://doi.org/10.1038/nature26160} {\bibfield
  {journal} {\bibinfo  {journal} {Nature}\ }\textbf {\bibinfo {volume} {556}},\
  \bibinfo {pages} {43} (\bibinfo {year} {2018}{\natexlab{a}})}\BibitemShut
  {NoStop}%
\bibitem [{\citenamefont {Cao}\ \emph {et~al.}(2018{\natexlab{b}})\citenamefont
  {Cao}, \citenamefont {Fatemi}, \citenamefont {Demir}, \citenamefont {Fang},
  \citenamefont {Tomarken}, \citenamefont {Luo}, \citenamefont
  {Sanchez-Yamagishi}, \citenamefont {Watanabe}, \citenamefont {Taniguchi},
  \citenamefont {Kaxiras}, \citenamefont {Ashoori},\ and\ \citenamefont
  {Jarillo-Herrero}}]{cao2018insulator}%
  \BibitemOpen
  \bibfield  {author} {\bibinfo {author} {\bibfnamefont {Y.}~\bibnamefont
  {Cao}}, \bibinfo {author} {\bibfnamefont {V.}~\bibnamefont {Fatemi}},
  \bibinfo {author} {\bibfnamefont {A.}~\bibnamefont {Demir}}, \bibinfo
  {author} {\bibfnamefont {S.}~\bibnamefont {Fang}}, \bibinfo {author}
  {\bibfnamefont {S.~L.}\ \bibnamefont {Tomarken}}, \bibinfo {author}
  {\bibfnamefont {J.~Y.}\ \bibnamefont {Luo}}, \bibinfo {author} {\bibfnamefont
  {J.~D.}\ \bibnamefont {Sanchez-Yamagishi}}, \bibinfo {author} {\bibfnamefont
  {K.}~\bibnamefont {Watanabe}}, \bibinfo {author} {\bibfnamefont
  {T.}~\bibnamefont {Taniguchi}}, \bibinfo {author} {\bibfnamefont
  {E.}~\bibnamefont {Kaxiras}}, \bibinfo {author} {\bibfnamefont {R.~C.}\
  \bibnamefont {Ashoori}},\ and\ \bibinfo {author} {\bibfnamefont
  {P.}~\bibnamefont {Jarillo-Herrero}},\ }\bibfield  {title} {\bibinfo {title}
  {Correlated insulator behaviour at half-filling in magic-angle graphene
  superlattices},\ }\href {https://doi.org/10.1038/nature26154} {\bibfield
  {journal} {\bibinfo  {journal} {Nature}\ }\textbf {\bibinfo {volume} {556}},\
  \bibinfo {pages} {80} (\bibinfo {year} {2018}{\natexlab{b}})}\BibitemShut
  {NoStop}%
\bibitem [{\citenamefont {Yankowitz}\ \emph {et~al.}(2019)\citenamefont
  {Yankowitz}, \citenamefont {Chen}, \citenamefont {Polshyn}, \citenamefont
  {Zhang}, \citenamefont {Watanabe}, \citenamefont {Taniguchi}, \citenamefont
  {Graf}, \citenamefont {Young},\ and\ \citenamefont {Dean}}]{yankowitz2019}%
  \BibitemOpen
  \bibfield  {author} {\bibinfo {author} {\bibfnamefont {M.}~\bibnamefont
  {Yankowitz}}, \bibinfo {author} {\bibfnamefont {S.}~\bibnamefont {Chen}},
  \bibinfo {author} {\bibfnamefont {H.}~\bibnamefont {Polshyn}}, \bibinfo
  {author} {\bibfnamefont {Y.}~\bibnamefont {Zhang}}, \bibinfo {author}
  {\bibfnamefont {K.}~\bibnamefont {Watanabe}}, \bibinfo {author}
  {\bibfnamefont {T.}~\bibnamefont {Taniguchi}}, \bibinfo {author}
  {\bibfnamefont {D.}~\bibnamefont {Graf}}, \bibinfo {author} {\bibfnamefont
  {A.~F.}\ \bibnamefont {Young}},\ and\ \bibinfo {author} {\bibfnamefont
  {C.~R.}\ \bibnamefont {Dean}},\ }\bibfield  {title} {\bibinfo {title} {Tuning
  superconductivity in twisted bilayer graphene},\ }\href
  {https://doi.org/10.1126/science.aav1910} {\bibfield  {journal} {\bibinfo
  {journal} {Science}\ }\textbf {\bibinfo {volume} {363}},\ \bibinfo {pages}
  {1059} (\bibinfo {year} {2019})},\ \Eprint
  {https://arxiv.org/abs/https://www.science.org/doi/pdf/10.1126/science.aav1910}
  {https://www.science.org/doi/pdf/10.1126/science.aav1910} \BibitemShut
  {NoStop}%
\bibitem [{\citenamefont {Lu}\ \emph {et~al.}(2019)\citenamefont {Lu},
  \citenamefont {Stepanov}, \citenamefont {Yang}, \citenamefont {Xie},
  \citenamefont {Aamir}, \citenamefont {Das}, \citenamefont {Urgell},
  \citenamefont {Watanabe}, \citenamefont {Taniguchi}, \citenamefont {Zhang},
  \citenamefont {Bachtold}, \citenamefont {MacDonald},\ and\ \citenamefont
  {Efetov}}]{lu2019}%
  \BibitemOpen
  \bibfield  {author} {\bibinfo {author} {\bibfnamefont {X.}~\bibnamefont
  {Lu}}, \bibinfo {author} {\bibfnamefont {P.}~\bibnamefont {Stepanov}},
  \bibinfo {author} {\bibfnamefont {W.}~\bibnamefont {Yang}}, \bibinfo {author}
  {\bibfnamefont {M.}~\bibnamefont {Xie}}, \bibinfo {author} {\bibfnamefont
  {M.~A.}\ \bibnamefont {Aamir}}, \bibinfo {author} {\bibfnamefont
  {I.}~\bibnamefont {Das}}, \bibinfo {author} {\bibfnamefont {C.}~\bibnamefont
  {Urgell}}, \bibinfo {author} {\bibfnamefont {K.}~\bibnamefont {Watanabe}},
  \bibinfo {author} {\bibfnamefont {T.}~\bibnamefont {Taniguchi}}, \bibinfo
  {author} {\bibfnamefont {G.}~\bibnamefont {Zhang}}, \bibinfo {author}
  {\bibfnamefont {A.}~\bibnamefont {Bachtold}}, \bibinfo {author}
  {\bibfnamefont {A.~H.}\ \bibnamefont {MacDonald}},\ and\ \bibinfo {author}
  {\bibfnamefont {D.~K.}\ \bibnamefont {Efetov}},\ }\bibfield  {title}
  {\bibinfo {title} {Superconductors, orbital magnets and correlated states in
  magic-angle bilayer graphene},\ }\href
  {https://doi.org/10.1038/s41586-019-1695-0} {\bibfield  {journal} {\bibinfo
  {journal} {Nature}\ }\textbf {\bibinfo {volume} {574}},\ \bibinfo {pages}
  {653} (\bibinfo {year} {2019})}\BibitemShut {NoStop}%
\bibitem [{\citenamefont {Andrei}\ and\ \citenamefont
  {MacDonald}(2020)}]{andrei2020}%
  \BibitemOpen
  \bibfield  {author} {\bibinfo {author} {\bibfnamefont {E.~Y.}\ \bibnamefont
  {Andrei}}\ and\ \bibinfo {author} {\bibfnamefont {A.~H.}\ \bibnamefont
  {MacDonald}},\ }\bibfield  {title} {\bibinfo {title} {Graphene bilayers with
  a twist},\ }\href {https://doi.org/10.1038/s41563-020-00840-0} {\bibfield
  {journal} {\bibinfo  {journal} {Nature Materials}\ }\textbf {\bibinfo
  {volume} {19}},\ \bibinfo {pages} {1265} (\bibinfo {year}
  {2020})}\BibitemShut {NoStop}%
\bibitem [{\citenamefont {Cao}\ \emph {et~al.}(2021)\citenamefont {Cao},
  \citenamefont {Rodan-Legrain}, \citenamefont {Park}, \citenamefont {Yuan},
  \citenamefont {Watanabe}, \citenamefont {Taniguchi}, \citenamefont
  {Fernandes}, \citenamefont {Fu},\ and\ \citenamefont
  {Jarillo-Herrero}}]{cao2021}%
  \BibitemOpen
  \bibfield  {author} {\bibinfo {author} {\bibfnamefont {Y.}~\bibnamefont
  {Cao}}, \bibinfo {author} {\bibfnamefont {D.}~\bibnamefont {Rodan-Legrain}},
  \bibinfo {author} {\bibfnamefont {J.~M.}\ \bibnamefont {Park}}, \bibinfo
  {author} {\bibfnamefont {N.~F.~Q.}\ \bibnamefont {Yuan}}, \bibinfo {author}
  {\bibfnamefont {K.}~\bibnamefont {Watanabe}}, \bibinfo {author}
  {\bibfnamefont {T.}~\bibnamefont {Taniguchi}}, \bibinfo {author}
  {\bibfnamefont {R.~M.}\ \bibnamefont {Fernandes}}, \bibinfo {author}
  {\bibfnamefont {L.}~\bibnamefont {Fu}},\ and\ \bibinfo {author}
  {\bibfnamefont {P.}~\bibnamefont {Jarillo-Herrero}},\ }\bibfield  {title}
  {\bibinfo {title} {Nematicity and competing orders in superconducting
  magic-angle graphene},\ }\href {https://doi.org/10.1126/science.abc2836}
  {\bibfield  {journal} {\bibinfo  {journal} {Science}\ }\textbf {\bibinfo
  {volume} {372}},\ \bibinfo {pages} {264} (\bibinfo {year} {2021})},\ \Eprint
  {https://arxiv.org/abs/https://www.science.org/doi/pdf/10.1126/science.abc2836}
  {https://www.science.org/doi/pdf/10.1126/science.abc2836} \BibitemShut
  {NoStop}%
\bibitem [{\citenamefont {Zhou}\ \emph
  {et~al.}(2021{\natexlab{a}})\citenamefont {Zhou}, \citenamefont {Xie},
  \citenamefont {Ghazaryan}, \citenamefont {Holder}, \citenamefont {Ehrets},
  \citenamefont {Spanton}, \citenamefont {Taniguchi}, \citenamefont {Watanabe},
  \citenamefont {Berg}, \citenamefont {Serbyn},\ and\ \citenamefont
  {Young}}]{zhou2021rtg}%
  \BibitemOpen
  \bibfield  {author} {\bibinfo {author} {\bibfnamefont {H.}~\bibnamefont
  {Zhou}}, \bibinfo {author} {\bibfnamefont {T.}~\bibnamefont {Xie}}, \bibinfo
  {author} {\bibfnamefont {A.}~\bibnamefont {Ghazaryan}}, \bibinfo {author}
  {\bibfnamefont {T.}~\bibnamefont {Holder}}, \bibinfo {author} {\bibfnamefont
  {J.~R.}\ \bibnamefont {Ehrets}}, \bibinfo {author} {\bibfnamefont {E.~M.}\
  \bibnamefont {Spanton}}, \bibinfo {author} {\bibfnamefont {T.}~\bibnamefont
  {Taniguchi}}, \bibinfo {author} {\bibfnamefont {K.}~\bibnamefont {Watanabe}},
  \bibinfo {author} {\bibfnamefont {E.}~\bibnamefont {Berg}}, \bibinfo {author}
  {\bibfnamefont {M.}~\bibnamefont {Serbyn}},\ and\ \bibinfo {author}
  {\bibfnamefont {A.~F.}\ \bibnamefont {Young}},\ }\bibfield  {title} {\bibinfo
  {title} {Half- and quarter-metals in rhombohedral trilayer graphene},\ }\href
  {https://doi.org/10.1038/s41586-021-03938-w} {\bibfield  {journal} {\bibinfo
  {journal} {Nature}\ }\textbf {\bibinfo {volume} {598}},\ \bibinfo {pages}
  {429} (\bibinfo {year} {2021}{\natexlab{a}})}\BibitemShut {NoStop}%
\bibitem [{\citenamefont {Zhou}\ \emph
  {et~al.}(2021{\natexlab{b}})\citenamefont {Zhou}, \citenamefont {Xie},
  \citenamefont {Taniguchi}, \citenamefont {Watanabe},\ and\ \citenamefont
  {Young}}]{zhou2021sc}%
  \BibitemOpen
  \bibfield  {author} {\bibinfo {author} {\bibfnamefont {H.}~\bibnamefont
  {Zhou}}, \bibinfo {author} {\bibfnamefont {T.}~\bibnamefont {Xie}}, \bibinfo
  {author} {\bibfnamefont {T.}~\bibnamefont {Taniguchi}}, \bibinfo {author}
  {\bibfnamefont {K.}~\bibnamefont {Watanabe}},\ and\ \bibinfo {author}
  {\bibfnamefont {A.~F.}\ \bibnamefont {Young}},\ }\bibfield  {title} {\bibinfo
  {title} {Superconductivity in rhombohedral trilayer graphene},\ }\href
  {https://doi.org/10.1038/s41586-021-03926-0} {\bibfield  {journal} {\bibinfo
  {journal} {Nature}\ }\textbf {\bibinfo {volume} {598}},\ \bibinfo {pages}
  {434} (\bibinfo {year} {2021}{\natexlab{b}})}\BibitemShut {NoStop}%
\bibitem [{\citenamefont {Zhou}\ \emph {et~al.}(2022)\citenamefont {Zhou},
  \citenamefont {Holleis}, \citenamefont {Saito}, \citenamefont {Cohen},
  \citenamefont {Huynh}, \citenamefont {Patterson}, \citenamefont {Yang},
  \citenamefont {Taniguchi}, \citenamefont {Watanabe},\ and\ \citenamefont
  {Young}}]{zhou2022bbg}%
  \BibitemOpen
  \bibfield  {author} {\bibinfo {author} {\bibfnamefont {H.}~\bibnamefont
  {Zhou}}, \bibinfo {author} {\bibfnamefont {L.}~\bibnamefont {Holleis}},
  \bibinfo {author} {\bibfnamefont {Y.}~\bibnamefont {Saito}}, \bibinfo
  {author} {\bibfnamefont {L.}~\bibnamefont {Cohen}}, \bibinfo {author}
  {\bibfnamefont {W.}~\bibnamefont {Huynh}}, \bibinfo {author} {\bibfnamefont
  {C.~L.}\ \bibnamefont {Patterson}}, \bibinfo {author} {\bibfnamefont
  {F.}~\bibnamefont {Yang}}, \bibinfo {author} {\bibfnamefont {T.}~\bibnamefont
  {Taniguchi}}, \bibinfo {author} {\bibfnamefont {K.}~\bibnamefont
  {Watanabe}},\ and\ \bibinfo {author} {\bibfnamefont {A.~F.}\ \bibnamefont
  {Young}},\ }\bibfield  {title} {\bibinfo {title} {Isospin magnetism and
  spin-polarized superconductivity in bernal bilayer graphene},\ }\href
  {https://doi.org/10.1126/science.abm8386} {\bibfield  {journal} {\bibinfo
  {journal} {Science}\ }\textbf {\bibinfo {volume} {375}},\ \bibinfo {pages}
  {774} (\bibinfo {year} {2022})},\ \Eprint
  {https://arxiv.org/abs/https://www.science.org/doi/pdf/10.1126/science.abm8386}
  {https://www.science.org/doi/pdf/10.1126/science.abm8386} \BibitemShut
  {NoStop}%
\bibitem [{\citenamefont {Seiler}\ \emph {et~al.}(2022)\citenamefont {Seiler},
  \citenamefont {Geisenhof}, \citenamefont {Winterer}, \citenamefont
  {Watanabe}, \citenamefont {Taniguchi}, \citenamefont {Xu}, \citenamefont
  {Zhang},\ and\ \citenamefont {Weitz}}]{seiler2022}%
  \BibitemOpen
  \bibfield  {author} {\bibinfo {author} {\bibfnamefont {A.~M.}\ \bibnamefont
  {Seiler}}, \bibinfo {author} {\bibfnamefont {F.~R.}\ \bibnamefont
  {Geisenhof}}, \bibinfo {author} {\bibfnamefont {F.}~\bibnamefont {Winterer}},
  \bibinfo {author} {\bibfnamefont {K.}~\bibnamefont {Watanabe}}, \bibinfo
  {author} {\bibfnamefont {T.}~\bibnamefont {Taniguchi}}, \bibinfo {author}
  {\bibfnamefont {T.}~\bibnamefont {Xu}}, \bibinfo {author} {\bibfnamefont
  {F.}~\bibnamefont {Zhang}},\ and\ \bibinfo {author} {\bibfnamefont {R.~T.}\
  \bibnamefont {Weitz}},\ }\bibfield  {title} {\bibinfo {title} {Quantum
  cascade of correlated phases in trigonally warped bilayer graphene},\ }\href
  {https://doi.org/10.1038/s41586-022-04937-1} {\bibfield  {journal} {\bibinfo
  {journal} {Nature}\ }\textbf {\bibinfo {volume} {608}},\ \bibinfo {pages}
  {298} (\bibinfo {year} {2022})}\BibitemShut {NoStop}%
\bibitem [{\citenamefont {Seiler}\ \emph {et~al.}(2025)\citenamefont {Seiler},
  \citenamefont {Zhumagulov}, \citenamefont {Zollner}, \citenamefont {Yoon},
  \citenamefont {Urbaniak}, \citenamefont {Geisenhof}, \citenamefont
  {Watanabe}, \citenamefont {Taniguchi}, \citenamefont {Fabian}, \citenamefont
  {Zhang},\ and\ \citenamefont {Weitz}}]{Seiler2024}%
  \BibitemOpen
  \bibfield  {author} {\bibinfo {author} {\bibfnamefont {A.~M.}\ \bibnamefont
  {Seiler}}, \bibinfo {author} {\bibfnamefont {Y.}~\bibnamefont {Zhumagulov}},
  \bibinfo {author} {\bibfnamefont {K.}~\bibnamefont {Zollner}}, \bibinfo
  {author} {\bibfnamefont {C.}~\bibnamefont {Yoon}}, \bibinfo {author}
  {\bibfnamefont {D.}~\bibnamefont {Urbaniak}}, \bibinfo {author}
  {\bibfnamefont {F.~R.}\ \bibnamefont {Geisenhof}}, \bibinfo {author}
  {\bibfnamefont {K.}~\bibnamefont {Watanabe}}, \bibinfo {author}
  {\bibfnamefont {T.}~\bibnamefont {Taniguchi}}, \bibinfo {author}
  {\bibfnamefont {J.}~\bibnamefont {Fabian}}, \bibinfo {author} {\bibfnamefont
  {F.}~\bibnamefont {Zhang}},\ and\ \bibinfo {author} {\bibfnamefont {R.~T.}\
  \bibnamefont {Weitz}},\ }\bibfield  {title} {\bibinfo {title}
  {Layer-selective spin-orbit coupling and strong correlation in bilayer
  graphene},\ }\href {https://doi.org/10.1088/2053-1583/add74a} {\bibfield
  {journal} {\bibinfo  {journal} {2D Materials}\ }\textbf {\bibinfo {volume}
  {12}},\ \bibinfo {pages} {035009} (\bibinfo {year} {2025})}\BibitemShut
  {NoStop}%
\bibitem [{\citenamefont {de~la Barrera}\ \emph {et~al.}(2022)\citenamefont
  {de~la Barrera}, \citenamefont {Aronson}, \citenamefont {Zheng},
  \citenamefont {Watanabe}, \citenamefont {Taniguchi}, \citenamefont {Ma},
  \citenamefont {Jarillo-Herrero},\ and\ \citenamefont
  {Ashoori}}]{barrera2022}%
  \BibitemOpen
  \bibfield  {author} {\bibinfo {author} {\bibfnamefont {S.~C.}\ \bibnamefont
  {de~la Barrera}}, \bibinfo {author} {\bibfnamefont {S.}~\bibnamefont
  {Aronson}}, \bibinfo {author} {\bibfnamefont {Z.}~\bibnamefont {Zheng}},
  \bibinfo {author} {\bibfnamefont {K.}~\bibnamefont {Watanabe}}, \bibinfo
  {author} {\bibfnamefont {T.}~\bibnamefont {Taniguchi}}, \bibinfo {author}
  {\bibfnamefont {Q.}~\bibnamefont {Ma}}, \bibinfo {author} {\bibfnamefont
  {P.}~\bibnamefont {Jarillo-Herrero}},\ and\ \bibinfo {author} {\bibfnamefont
  {R.}~\bibnamefont {Ashoori}},\ }\bibfield  {title} {\bibinfo {title} {Cascade
  of isospin phase transitions in bernal-stacked bilayer graphene at zero
  magnetic field},\ }\href {https://doi.org/10.1038/s41567-022-01616-w}
  {\bibfield  {journal} {\bibinfo  {journal} {Nature Physics}\ }\textbf
  {\bibinfo {volume} {18}},\ \bibinfo {pages} {771} (\bibinfo {year}
  {2022})}\BibitemShut {NoStop}%
\bibitem [{\citenamefont {Han}\ \emph {et~al.}(2023)\citenamefont {Han},
  \citenamefont {Lu}, \citenamefont {Scuri}, \citenamefont {Sung},
  \citenamefont {Wang}, \citenamefont {Han}, \citenamefont {Watanabe},
  \citenamefont {Taniguchi}, \citenamefont {Fu}, \citenamefont {Park},\ and\
  \citenamefont {Ju}}]{han2023}%
  \BibitemOpen
  \bibfield  {author} {\bibinfo {author} {\bibfnamefont {T.}~\bibnamefont
  {Han}}, \bibinfo {author} {\bibfnamefont {Z.}~\bibnamefont {Lu}}, \bibinfo
  {author} {\bibfnamefont {G.}~\bibnamefont {Scuri}}, \bibinfo {author}
  {\bibfnamefont {J.}~\bibnamefont {Sung}}, \bibinfo {author} {\bibfnamefont
  {J.}~\bibnamefont {Wang}}, \bibinfo {author} {\bibfnamefont {T.}~\bibnamefont
  {Han}}, \bibinfo {author} {\bibfnamefont {K.}~\bibnamefont {Watanabe}},
  \bibinfo {author} {\bibfnamefont {T.}~\bibnamefont {Taniguchi}}, \bibinfo
  {author} {\bibfnamefont {L.}~\bibnamefont {Fu}}, \bibinfo {author}
  {\bibfnamefont {H.}~\bibnamefont {Park}},\ and\ \bibinfo {author}
  {\bibfnamefont {L.}~\bibnamefont {Ju}},\ }\bibfield  {title} {\bibinfo
  {title} {Orbital multiferroicity in pentalayer rhombohedral graphene},\
  }\href {https://doi.org/10.1038/s41586-023-06572-w} {\bibfield  {journal}
  {\bibinfo  {journal} {Nature}\ }\textbf {\bibinfo {volume} {623}},\ \bibinfo
  {pages} {41} (\bibinfo {year} {2023})}\BibitemShut {NoStop}%
\bibitem [{\citenamefont {Han}\ \emph {et~al.}(2024)\citenamefont {Han},
  \citenamefont {Lu}, \citenamefont {Scuri}, \citenamefont {Sung},
  \citenamefont {Wang}, \citenamefont {Han}, \citenamefont {Watanabe},
  \citenamefont {Taniguchi}, \citenamefont {Park},\ and\ \citenamefont
  {Ju}}]{han2024}%
  \BibitemOpen
  \bibfield  {author} {\bibinfo {author} {\bibfnamefont {T.}~\bibnamefont
  {Han}}, \bibinfo {author} {\bibfnamefont {Z.}~\bibnamefont {Lu}}, \bibinfo
  {author} {\bibfnamefont {G.}~\bibnamefont {Scuri}}, \bibinfo {author}
  {\bibfnamefont {J.}~\bibnamefont {Sung}}, \bibinfo {author} {\bibfnamefont
  {J.}~\bibnamefont {Wang}}, \bibinfo {author} {\bibfnamefont {T.}~\bibnamefont
  {Han}}, \bibinfo {author} {\bibfnamefont {K.}~\bibnamefont {Watanabe}},
  \bibinfo {author} {\bibfnamefont {T.}~\bibnamefont {Taniguchi}}, \bibinfo
  {author} {\bibfnamefont {H.}~\bibnamefont {Park}},\ and\ \bibinfo {author}
  {\bibfnamefont {L.}~\bibnamefont {Ju}},\ }\bibfield  {title} {\bibinfo
  {title} {Correlated insulator and chern insulators in pentalayer
  rhombohedral-stacked graphene},\ }\href
  {https://doi.org/10.1038/s41565-023-01520-1} {\bibfield  {journal} {\bibinfo
  {journal} {Nature Nanotechnology}\ }\textbf {\bibinfo {volume} {19}},\
  \bibinfo {pages} {181} (\bibinfo {year} {2024})}\BibitemShut {NoStop}%
\bibitem [{\citenamefont {Arp}\ \emph {et~al.}(2024)\citenamefont {Arp},
  \citenamefont {Sheekey}, \citenamefont {Zhou}, \citenamefont {Tschirhart},
  \citenamefont {Patterson}, \citenamefont {Yoo}, \citenamefont {Holleis},
  \citenamefont {Redekop}, \citenamefont {Babikyan}, \citenamefont {Xie},
  \citenamefont {Xiao}, \citenamefont {Vituri}, \citenamefont {Holder},
  \citenamefont {Taniguchi}, \citenamefont {Watanabe}, \citenamefont {Huber},
  \citenamefont {Berg},\ and\ \citenamefont {Young}}]{Arp2024}%
  \BibitemOpen
  \bibfield  {author} {\bibinfo {author} {\bibfnamefont {T.}~\bibnamefont
  {Arp}}, \bibinfo {author} {\bibfnamefont {O.}~\bibnamefont {Sheekey}},
  \bibinfo {author} {\bibfnamefont {H.}~\bibnamefont {Zhou}}, \bibinfo {author}
  {\bibfnamefont {C.~L.}\ \bibnamefont {Tschirhart}}, \bibinfo {author}
  {\bibfnamefont {C.~L.}\ \bibnamefont {Patterson}}, \bibinfo {author}
  {\bibfnamefont {H.~M.}\ \bibnamefont {Yoo}}, \bibinfo {author} {\bibfnamefont
  {L.}~\bibnamefont {Holleis}}, \bibinfo {author} {\bibfnamefont
  {E.}~\bibnamefont {Redekop}}, \bibinfo {author} {\bibfnamefont
  {G.}~\bibnamefont {Babikyan}}, \bibinfo {author} {\bibfnamefont
  {T.}~\bibnamefont {Xie}}, \bibinfo {author} {\bibfnamefont {J.}~\bibnamefont
  {Xiao}}, \bibinfo {author} {\bibfnamefont {Y.}~\bibnamefont {Vituri}},
  \bibinfo {author} {\bibfnamefont {T.}~\bibnamefont {Holder}}, \bibinfo
  {author} {\bibfnamefont {T.}~\bibnamefont {Taniguchi}}, \bibinfo {author}
  {\bibfnamefont {K.}~\bibnamefont {Watanabe}}, \bibinfo {author}
  {\bibfnamefont {M.~E.}\ \bibnamefont {Huber}}, \bibinfo {author}
  {\bibfnamefont {E.}~\bibnamefont {Berg}},\ and\ \bibinfo {author}
  {\bibfnamefont {A.~F.}\ \bibnamefont {Young}},\ }\bibfield  {title} {\bibinfo
  {title} {Intervalley coherence and intrinsic spin--orbit coupling in
  rhombohedral trilayer graphene},\ }\href
  {https://doi.org/10.1038/s41567-024-02560-7} {\bibfield  {journal} {\bibinfo
  {journal} {Nature Physics}\ }\textbf {\bibinfo {volume} {20}},\ \bibinfo
  {pages} {1413} (\bibinfo {year} {2024})}\BibitemShut {NoStop}%
\bibitem [{\citenamefont {Holleis}\ \emph {et~al.}(2025)\citenamefont
  {Holleis}, \citenamefont {Patterson}, \citenamefont {Zhang}, \citenamefont
  {Vituri}, \citenamefont {Yoo}, \citenamefont {Zhou}, \citenamefont
  {Taniguchi}, \citenamefont {Watanabe}, \citenamefont {Berg}, \citenamefont
  {Nadj-Perge},\ and\ \citenamefont {Young}}]{holleis2025}%
  \BibitemOpen
  \bibfield  {author} {\bibinfo {author} {\bibfnamefont {L.}~\bibnamefont
  {Holleis}}, \bibinfo {author} {\bibfnamefont {C.~L.}\ \bibnamefont
  {Patterson}}, \bibinfo {author} {\bibfnamefont {Y.}~\bibnamefont {Zhang}},
  \bibinfo {author} {\bibfnamefont {Y.}~\bibnamefont {Vituri}}, \bibinfo
  {author} {\bibfnamefont {H.~M.}\ \bibnamefont {Yoo}}, \bibinfo {author}
  {\bibfnamefont {H.}~\bibnamefont {Zhou}}, \bibinfo {author} {\bibfnamefont
  {T.}~\bibnamefont {Taniguchi}}, \bibinfo {author} {\bibfnamefont
  {K.}~\bibnamefont {Watanabe}}, \bibinfo {author} {\bibfnamefont
  {E.}~\bibnamefont {Berg}}, \bibinfo {author} {\bibfnamefont {S.}~\bibnamefont
  {Nadj-Perge}},\ and\ \bibinfo {author} {\bibfnamefont {A.~F.}\ \bibnamefont
  {Young}},\ }\bibfield  {title} {\bibinfo {title} {Nematicity and orbital
  depairing in superconducting bernal bilayer graphene},\ }\href
  {https://doi.org/10.1038/s41567-024-02776-7} {\bibfield  {journal} {\bibinfo
  {journal} {Nature Physics}\ }\textbf {\bibinfo {volume} {21}},\ \bibinfo
  {pages} {444} (\bibinfo {year} {2025})}\BibitemShut {NoStop}%
\bibitem [{\citenamefont {Chichinadze}\ \emph
  {et~al.}(2022{\natexlab{a}})\citenamefont {Chichinadze}, \citenamefont
  {Classen}, \citenamefont {Wang},\ and\ \citenamefont
  {Chubukov}}]{chichinadze2022}%
  \BibitemOpen
  \bibfield  {author} {\bibinfo {author} {\bibfnamefont {D.~V.}\ \bibnamefont
  {Chichinadze}}, \bibinfo {author} {\bibfnamefont {L.}~\bibnamefont
  {Classen}}, \bibinfo {author} {\bibfnamefont {Y.}~\bibnamefont {Wang}},\ and\
  \bibinfo {author} {\bibfnamefont {A.~V.}\ \bibnamefont {Chubukov}},\
  }\bibfield  {title} {\bibinfo {title} {Cascade of transitions in twisted and
  non-twisted graphene layers within the van hove scenario},\ }\href
  {https://doi.org/10.1038/s41535-022-00520-z} {\bibfield  {journal} {\bibinfo
  {journal} {npj Quantum Materials}\ }\textbf {\bibinfo {volume} {7}},\
  \bibinfo {pages} {114} (\bibinfo {year} {2022}{\natexlab{a}})}\BibitemShut
  {NoStop}%
\bibitem [{\citenamefont {Chichinadze}\ \emph
  {et~al.}(2022{\natexlab{b}})\citenamefont {Chichinadze}, \citenamefont
  {Classen}, \citenamefont {Wang},\ and\ \citenamefont
  {Chubukov}}]{chichinadze2022letters}%
  \BibitemOpen
  \bibfield  {author} {\bibinfo {author} {\bibfnamefont {D.~V.}\ \bibnamefont
  {Chichinadze}}, \bibinfo {author} {\bibfnamefont {L.}~\bibnamefont
  {Classen}}, \bibinfo {author} {\bibfnamefont {Y.}~\bibnamefont {Wang}},\ and\
  \bibinfo {author} {\bibfnamefont {A.~V.}\ \bibnamefont {Chubukov}},\
  }\bibfield  {title} {\bibinfo {title} {Su(4) symmetry in twisted bilayer
  graphene: An itinerant perspective},\ }\href
  {https://doi.org/10.1103/PhysRevLett.128.227601} {\bibfield  {journal}
  {\bibinfo  {journal} {Phys. Rev. Lett.}\ }\textbf {\bibinfo {volume} {128}},\
  \bibinfo {pages} {227601} (\bibinfo {year} {2022}{\natexlab{b}})}\BibitemShut
  {NoStop}%
\bibitem [{\citenamefont {Hu}\ \emph {et~al.}(2023)\citenamefont {Hu},
  \citenamefont {Bernevig},\ and\ \citenamefont {Tsvelik}}]{haoyu2023}%
  \BibitemOpen
  \bibfield  {author} {\bibinfo {author} {\bibfnamefont {H.}~\bibnamefont
  {Hu}}, \bibinfo {author} {\bibfnamefont {B.~A.}\ \bibnamefont {Bernevig}},\
  and\ \bibinfo {author} {\bibfnamefont {A.~M.}\ \bibnamefont {Tsvelik}},\
  }\bibfield  {title} {\bibinfo {title} {Kondo lattice model of magic-angle
  twisted-bilayer graphene: Hund's rule, local-moment fluctuations, and
  low-energy effective theory},\ }\href
  {https://doi.org/10.1103/PhysRevLett.131.026502} {\bibfield  {journal}
  {\bibinfo  {journal} {Phys. Rev. Lett.}\ }\textbf {\bibinfo {volume} {131}},\
  \bibinfo {pages} {026502} (\bibinfo {year} {2023})}\BibitemShut {NoStop}%
\bibitem [{\citenamefont {Xie}\ and\ \citenamefont
  {Das~Sarma}(2023)}]{xie2023}%
  \BibitemOpen
  \bibfield  {author} {\bibinfo {author} {\bibfnamefont {M.}~\bibnamefont
  {Xie}}\ and\ \bibinfo {author} {\bibfnamefont {S.}~\bibnamefont
  {Das~Sarma}},\ }\bibfield  {title} {\bibinfo {title} {Flavor symmetry
  breaking in spin-orbit coupled bilayer graphene},\ }\href
  {https://doi.org/10.1103/PhysRevB.107.L201119} {\bibfield  {journal}
  {\bibinfo  {journal} {Phys. Rev. B}\ }\textbf {\bibinfo {volume} {107}},\
  \bibinfo {pages} {L201119} (\bibinfo {year} {2023})}\BibitemShut {NoStop}%
\bibitem [{\citenamefont {Lee}\ \emph {et~al.}(2024)\citenamefont {Lee},
  \citenamefont {Chichinadze},\ and\ \citenamefont {Chubukov}}]{lee2024}%
  \BibitemOpen
  \bibfield  {author} {\bibinfo {author} {\bibfnamefont {Y.-C.}\ \bibnamefont
  {Lee}}, \bibinfo {author} {\bibfnamefont {D.~V.}\ \bibnamefont
  {Chichinadze}},\ and\ \bibinfo {author} {\bibfnamefont {A.~V.}\ \bibnamefont
  {Chubukov}},\ }\bibfield  {title} {\bibinfo {title} {Crossover from ordinary
  to higher order van hove singularity in a honeycomb system: A parquet
  renormalization group analysis},\ }\href
  {https://doi.org/10.1103/PhysRevB.109.155118} {\bibfield  {journal} {\bibinfo
   {journal} {Phys. Rev. B}\ }\textbf {\bibinfo {volume} {109}},\ \bibinfo
  {pages} {155118} (\bibinfo {year} {2024})}\BibitemShut {NoStop}%
\bibitem [{\citenamefont {Koh}\ \emph {et~al.}(2024{\natexlab{a}})\citenamefont
  {Koh}, \citenamefont {Alicea},\ and\ \citenamefont
  {Lantagne-Hurtubise}}]{koh2024rtg}%
  \BibitemOpen
  \bibfield  {author} {\bibinfo {author} {\bibfnamefont {J.~M.}\ \bibnamefont
  {Koh}}, \bibinfo {author} {\bibfnamefont {J.}~\bibnamefont {Alicea}},\ and\
  \bibinfo {author} {\bibfnamefont {E.}~\bibnamefont {Lantagne-Hurtubise}},\
  }\bibfield  {title} {\bibinfo {title} {Correlated phases in
  spin-orbit-coupled rhombohedral trilayer graphene},\ }\href
  {https://doi.org/10.1103/PhysRevB.109.035113} {\bibfield  {journal} {\bibinfo
   {journal} {Phys. Rev. B}\ }\textbf {\bibinfo {volume} {109}},\ \bibinfo
  {pages} {035113} (\bibinfo {year} {2024}{\natexlab{a}})}\BibitemShut
  {NoStop}%
\bibitem [{\citenamefont {Koh}\ \emph {et~al.}(2024{\natexlab{b}})\citenamefont
  {Koh}, \citenamefont {Thomson}, \citenamefont {Alicea},\ and\ \citenamefont
  {Lantagne-Hurtubise}}]{koh2024bbg}%
  \BibitemOpen
  \bibfield  {author} {\bibinfo {author} {\bibfnamefont {J.~M.}\ \bibnamefont
  {Koh}}, \bibinfo {author} {\bibfnamefont {A.}~\bibnamefont {Thomson}},
  \bibinfo {author} {\bibfnamefont {J.}~\bibnamefont {Alicea}},\ and\ \bibinfo
  {author} {\bibfnamefont {E.}~\bibnamefont {Lantagne-Hurtubise}},\ }\bibfield
  {title} {\bibinfo {title} {Symmetry-broken metallic orders in
  spin-orbit-coupled bernal bilayer graphene},\ }\href
  {https://doi.org/10.1103/PhysRevB.110.245118} {\bibfield  {journal} {\bibinfo
   {journal} {Phys. Rev. B}\ }\textbf {\bibinfo {volume} {110}},\ \bibinfo
  {pages} {245118} (\bibinfo {year} {2024}{\natexlab{b}})}\BibitemShut
  {NoStop}%
\bibitem [{\citenamefont {Wang}\ \emph {et~al.}(2024)\citenamefont {Wang},
  \citenamefont {Vila}, \citenamefont {Zaletel},\ and\ \citenamefont
  {Chatterjee}}]{wang2024electrical}%
  \BibitemOpen
  \bibfield  {author} {\bibinfo {author} {\bibfnamefont {T.}~\bibnamefont
  {Wang}}, \bibinfo {author} {\bibfnamefont {M.}~\bibnamefont {Vila}}, \bibinfo
  {author} {\bibfnamefont {M.~P.}\ \bibnamefont {Zaletel}},\ and\ \bibinfo
  {author} {\bibfnamefont {S.}~\bibnamefont {Chatterjee}},\ }\bibfield  {title}
  {\bibinfo {title} {Electrical control of spin and valley in spin-orbit
  coupled graphene multilayers},\ }\href
  {https://doi.org/10.1103/PhysRevLett.132.116504} {\bibfield  {journal}
  {\bibinfo  {journal} {Phys. Rev. Lett.}\ }\textbf {\bibinfo {volume} {132}},\
  \bibinfo {pages} {116504} (\bibinfo {year} {2024})}\BibitemShut {NoStop}%
\bibitem [{\citenamefont {Friedlan}\ \emph {et~al.}(2025)\citenamefont
  {Friedlan}, \citenamefont {Li},\ and\ \citenamefont {Kee}}]{friedlan2025}%
  \BibitemOpen
  \bibfield  {author} {\bibinfo {author} {\bibfnamefont {A.}~\bibnamefont
  {Friedlan}}, \bibinfo {author} {\bibfnamefont {H.}~\bibnamefont {Li}},\ and\
  \bibinfo {author} {\bibfnamefont {H.-Y.}\ \bibnamefont {Kee}},\ }\bibfield
  {title} {\bibinfo {title} {Valley polarization, magnetization, and
  superconductivity in bilayer graphene near the van hove singularity},\ }\href
  {https://doi.org/10.1103/PhysRevB.111.024504} {\bibfield  {journal} {\bibinfo
   {journal} {Phys. Rev. B}\ }\textbf {\bibinfo {volume} {111}},\ \bibinfo
  {pages} {024504} (\bibinfo {year} {2025})}\BibitemShut {NoStop}%
\bibitem [{\citenamefont {Mayrhofer}\ and\ \citenamefont
  {Chubukov}(2025)}]{mayrhofer2025}%
  \BibitemOpen
  \bibfield  {author} {\bibinfo {author} {\bibfnamefont {R.~D.}\ \bibnamefont
  {Mayrhofer}}\ and\ \bibinfo {author} {\bibfnamefont {A.~V.}\ \bibnamefont
  {Chubukov}},\ }\bibfield  {title} {\bibinfo {title} {Valley- and
  spin-polarized states in bernal bilayer graphene},\ }\href
  {https://doi.org/10.1103/PhysRevB.111.245114} {\bibfield  {journal} {\bibinfo
   {journal} {Phys. Rev. B}\ }\textbf {\bibinfo {volume} {111}},\ \bibinfo
  {pages} {245114} (\bibinfo {year} {2025})}\BibitemShut {NoStop}%
\bibitem [{\citenamefont {Stoner}(1938)}]{stoner1938}%
  \BibitemOpen
  \bibfield  {author} {\bibinfo {author} {\bibfnamefont {E.~C.}\ \bibnamefont
  {Stoner}},\ }\bibfield  {title} {\bibinfo {title} {Collective electron
  ferromagnetism},\ }\href@noop {} {\bibfield  {journal} {\bibinfo  {journal}
  {Proc. R. Soc. Lond. A}\ }\textbf {\bibinfo {volume} {165}},\ \bibinfo
  {pages} {372} (\bibinfo {year} {1938})}\BibitemShut {NoStop}%
\bibitem [{\citenamefont {Shimizu}(1981)}]{Shimizu1981}%
  \BibitemOpen
  \bibfield  {author} {\bibinfo {author} {\bibfnamefont {M.}~\bibnamefont
  {Shimizu}},\ }\bibfield  {title} {\bibinfo {title} {Itinerant electron
  magnetism},\ }\href {https://doi.org/10.1088/0034-4885/44/4/001} {\bibfield
  {journal} {\bibinfo  {journal} {Reports on Progress in Physics}\ }\textbf
  {\bibinfo {volume} {44}},\ \bibinfo {pages} {329} (\bibinfo {year}
  {1981})}\BibitemShut {NoStop}%
\bibitem [{\citenamefont {Duine}\ and\ \citenamefont
  {MacDonald}(2005)}]{duine2005}%
  \BibitemOpen
  \bibfield  {author} {\bibinfo {author} {\bibfnamefont {R.~A.}\ \bibnamefont
  {Duine}}\ and\ \bibinfo {author} {\bibfnamefont {A.~H.}\ \bibnamefont
  {MacDonald}},\ }\bibfield  {title} {\bibinfo {title} {Itinerant
  ferromagnetism in an ultracold atom fermi gas},\ }\href
  {https://doi.org/10.1103/PhysRevLett.95.230403} {\bibfield  {journal}
  {\bibinfo  {journal} {Phys. Rev. Lett.}\ }\textbf {\bibinfo {volume} {95}},\
  \bibinfo {pages} {230403} (\bibinfo {year} {2005})}\BibitemShut {NoStop}%
\bibitem [{\citenamefont {He}\ \emph {et~al.}(2016)\citenamefont {He},
  \citenamefont {Liu}, \citenamefont {Huang},\ and\ \citenamefont
  {Hu}}]{he2016}%
  \BibitemOpen
  \bibfield  {author} {\bibinfo {author} {\bibfnamefont {L.}~\bibnamefont
  {He}}, \bibinfo {author} {\bibfnamefont {X.-J.}\ \bibnamefont {Liu}},
  \bibinfo {author} {\bibfnamefont {X.-G.}\ \bibnamefont {Huang}},\ and\
  \bibinfo {author} {\bibfnamefont {H.}~\bibnamefont {Hu}},\ }\bibfield
  {title} {\bibinfo {title} {Stoner ferromagnetism of a strongly interacting
  fermi gas in the quasirepulsive regime},\ }\href
  {https://doi.org/10.1103/PhysRevA.93.063629} {\bibfield  {journal} {\bibinfo
  {journal} {Phys. Rev. A}\ }\textbf {\bibinfo {volume} {93}},\ \bibinfo
  {pages} {063629} (\bibinfo {year} {2016})}\BibitemShut {NoStop}%
\bibitem [{\citenamefont {Zhu}\ \emph {et~al.}(2018)\citenamefont {Zhu},
  \citenamefont {Sheng}, \citenamefont {Fu},\ and\ \citenamefont
  {Sodemann}}]{zhu2018}%
  \BibitemOpen
  \bibfield  {author} {\bibinfo {author} {\bibfnamefont {Z.}~\bibnamefont
  {Zhu}}, \bibinfo {author} {\bibfnamefont {D.~N.}\ \bibnamefont {Sheng}},
  \bibinfo {author} {\bibfnamefont {L.}~\bibnamefont {Fu}},\ and\ \bibinfo
  {author} {\bibfnamefont {I.}~\bibnamefont {Sodemann}},\ }\bibfield  {title}
  {\bibinfo {title} {Valley stoner instability of the composite fermi sea},\
  }\href {https://doi.org/10.1103/PhysRevB.98.155104} {\bibfield  {journal}
  {\bibinfo  {journal} {Phys. Rev. B}\ }\textbf {\bibinfo {volume} {98}},\
  \bibinfo {pages} {155104} (\bibinfo {year} {2018})}\BibitemShut {NoStop}%
\bibitem [{\citenamefont {Valenti}\ \emph {et~al.}(2024)\citenamefont
  {Valenti}, \citenamefont {Calvera}, \citenamefont {Kivelson}, \citenamefont
  {Berg},\ and\ \citenamefont {Huber}}]{valenti2024}%
  \BibitemOpen
  \bibfield  {author} {\bibinfo {author} {\bibfnamefont {A.}~\bibnamefont
  {Valenti}}, \bibinfo {author} {\bibfnamefont {V.}~\bibnamefont {Calvera}},
  \bibinfo {author} {\bibfnamefont {S.~A.}\ \bibnamefont {Kivelson}}, \bibinfo
  {author} {\bibfnamefont {E.}~\bibnamefont {Berg}},\ and\ \bibinfo {author}
  {\bibfnamefont {S.~D.}\ \bibnamefont {Huber}},\ }\bibfield  {title} {\bibinfo
  {title} {Nematic metal in a multivalley electron gas: Variational monte carlo
  analysis and application to alas},\ }\href
  {https://doi.org/10.1103/PhysRevLett.132.266501} {\bibfield  {journal}
  {\bibinfo  {journal} {Phys. Rev. Lett.}\ }\textbf {\bibinfo {volume} {132}},\
  \bibinfo {pages} {266501} (\bibinfo {year} {2024})}\BibitemShut {NoStop}%
\bibitem [{\citenamefont {Calvera}\ \emph {et~al.}(2025)\citenamefont
  {Calvera}, \citenamefont {Valenti}, \citenamefont {Huber}, \citenamefont
  {Berg},\ and\ \citenamefont {Kivelson}}]{calvera2024nematicity}%
  \BibitemOpen
  \bibfield  {author} {\bibinfo {author} {\bibfnamefont {V.}~\bibnamefont
  {Calvera}}, \bibinfo {author} {\bibfnamefont {A.}~\bibnamefont {Valenti}},
  \bibinfo {author} {\bibfnamefont {S.~D.}\ \bibnamefont {Huber}}, \bibinfo
  {author} {\bibfnamefont {E.}~\bibnamefont {Berg}},\ and\ \bibinfo {author}
  {\bibfnamefont {S.~A.}\ \bibnamefont {Kivelson}},\ }\bibfield  {title}
  {\bibinfo {title} {Theory of coulomb driven nematicity in a multivalley
  two-dimensional electron gas},\ }\href
  {https://doi.org/10.1103/PhysRevB.111.155135} {\bibfield  {journal} {\bibinfo
   {journal} {Phys. Rev. B}\ }\textbf {\bibinfo {volume} {111}},\ \bibinfo
  {pages} {155135} (\bibinfo {year} {2025})}\BibitemShut {NoStop}%
\bibitem [{\citenamefont {Raines}\ and\ \citenamefont
  {Chubukov}(2024)}]{raines2024stoner}%
  \BibitemOpen
  \bibfield  {author} {\bibinfo {author} {\bibfnamefont {Z.~M.}\ \bibnamefont
  {Raines}}\ and\ \bibinfo {author} {\bibfnamefont {A.~V.}\ \bibnamefont
  {Chubukov}},\ }\bibfield  {title} {\bibinfo {title} {Two-dimensional stoner
  transitions beyond mean field},\ }\href
  {https://doi.org/10.1103/PhysRevB.110.235433} {\bibfield  {journal} {\bibinfo
   {journal} {Phys. Rev. B}\ }\textbf {\bibinfo {volume} {110}},\ \bibinfo
  {pages} {235433} (\bibinfo {year} {2024})}\BibitemShut {NoStop}%
\bibitem [{\citenamefont {Shayegan}\ \emph {et~al.}(2006)\citenamefont
  {Shayegan}, \citenamefont {De~Poortere}, \citenamefont {Gunawan},
  \citenamefont {Shkolnikov}, \citenamefont {Tutuc},\ and\ \citenamefont
  {Vakili}}]{shayegan2006}%
  \BibitemOpen
  \bibfield  {author} {\bibinfo {author} {\bibfnamefont {M.}~\bibnamefont
  {Shayegan}}, \bibinfo {author} {\bibfnamefont {E.~P.}\ \bibnamefont
  {De~Poortere}}, \bibinfo {author} {\bibfnamefont {O.}~\bibnamefont
  {Gunawan}}, \bibinfo {author} {\bibfnamefont {Y.~P.}\ \bibnamefont
  {Shkolnikov}}, \bibinfo {author} {\bibfnamefont {E.}~\bibnamefont {Tutuc}},\
  and\ \bibinfo {author} {\bibfnamefont {K.}~\bibnamefont {Vakili}},\
  }\bibfield  {title} {\bibinfo {title} {Two-dimensional electrons occupying
  multiple valleys in alas},\ }\href
  {https://doi.org/https://doi.org/10.1002/pssb.200642212} {\bibfield
  {journal} {\bibinfo  {journal} {physica status solidi (b)}\ }\textbf
  {\bibinfo {volume} {243}},\ \bibinfo {pages} {3629} (\bibinfo {year}
  {2006})},\ \Eprint
  {https://arxiv.org/abs/https://onlinelibrary.wiley.com/doi/pdf/10.1002/pssb.200642212}
  {https://onlinelibrary.wiley.com/doi/pdf/10.1002/pssb.200642212} \BibitemShut
  {NoStop}%
\bibitem [{\citenamefont {Gunawan}\ \emph {et~al.}(2006)\citenamefont
  {Gunawan}, \citenamefont {Shkolnikov}, \citenamefont {Vakili}, \citenamefont
  {Gokmen}, \citenamefont {De~Poortere},\ and\ \citenamefont
  {Shayegan}}]{gunawan2006}%
  \BibitemOpen
  \bibfield  {author} {\bibinfo {author} {\bibfnamefont {O.}~\bibnamefont
  {Gunawan}}, \bibinfo {author} {\bibfnamefont {Y.~P.}\ \bibnamefont
  {Shkolnikov}}, \bibinfo {author} {\bibfnamefont {K.}~\bibnamefont {Vakili}},
  \bibinfo {author} {\bibfnamefont {T.}~\bibnamefont {Gokmen}}, \bibinfo
  {author} {\bibfnamefont {E.~P.}\ \bibnamefont {De~Poortere}},\ and\ \bibinfo
  {author} {\bibfnamefont {M.}~\bibnamefont {Shayegan}},\ }\bibfield  {title}
  {\bibinfo {title} {Valley susceptibility of an interacting two-dimensional
  electron system},\ }\href {https://doi.org/10.1103/PhysRevLett.97.186404}
  {\bibfield  {journal} {\bibinfo  {journal} {Phys. Rev. Lett.}\ }\textbf
  {\bibinfo {volume} {97}},\ \bibinfo {pages} {186404} (\bibinfo {year}
  {2006})}\BibitemShut {NoStop}%
\bibitem [{\citenamefont {Hossain}\ \emph {et~al.}(2020)\citenamefont
  {Hossain}, \citenamefont {Ma}, \citenamefont {Rosales}, \citenamefont
  {Chung}, \citenamefont {Pfeiffer}, \citenamefont {West}, \citenamefont
  {Baldwin},\ and\ \citenamefont {Shayegan}}]{hossain2020}%
  \BibitemOpen
  \bibfield  {author} {\bibinfo {author} {\bibfnamefont {M.~S.}\ \bibnamefont
  {Hossain}}, \bibinfo {author} {\bibfnamefont {M.~K.}\ \bibnamefont {Ma}},
  \bibinfo {author} {\bibfnamefont {K.~A.~V.}\ \bibnamefont {Rosales}},
  \bibinfo {author} {\bibfnamefont {Y.~J.}\ \bibnamefont {Chung}}, \bibinfo
  {author} {\bibfnamefont {L.~N.}\ \bibnamefont {Pfeiffer}}, \bibinfo {author}
  {\bibfnamefont {K.~W.}\ \bibnamefont {West}}, \bibinfo {author}
  {\bibfnamefont {K.~W.}\ \bibnamefont {Baldwin}},\ and\ \bibinfo {author}
  {\bibfnamefont {M.}~\bibnamefont {Shayegan}},\ }\bibfield  {title} {\bibinfo
  {title} {Observation of spontaneous ferromagnetism in a two-dimensional
  electron system},\ }\href {https://doi.org/10.1073/pnas.2018248117}
  {\bibfield  {journal} {\bibinfo  {journal} {Proceedings of the National
  Academy of Sciences}\ }\textbf {\bibinfo {volume} {117}},\ \bibinfo {pages}
  {32244} (\bibinfo {year} {2020})},\ \Eprint
  {https://arxiv.org/abs/https://www.pnas.org/doi/pdf/10.1073/pnas.2018248117}
  {https://www.pnas.org/doi/pdf/10.1073/pnas.2018248117} \BibitemShut {NoStop}%
\bibitem [{\citenamefont {Hossain}\ \emph {et~al.}(2021)\citenamefont
  {Hossain}, \citenamefont {Ma}, \citenamefont {Villegas-Rosales},
  \citenamefont {Chung}, \citenamefont {Pfeiffer}, \citenamefont {West},
  \citenamefont {Baldwin},\ and\ \citenamefont {Shayegan}}]{hossain2021}%
  \BibitemOpen
  \bibfield  {author} {\bibinfo {author} {\bibfnamefont {M.~S.}\ \bibnamefont
  {Hossain}}, \bibinfo {author} {\bibfnamefont {M.~K.}\ \bibnamefont {Ma}},
  \bibinfo {author} {\bibfnamefont {K.~A.}\ \bibnamefont {Villegas-Rosales}},
  \bibinfo {author} {\bibfnamefont {Y.~J.}\ \bibnamefont {Chung}}, \bibinfo
  {author} {\bibfnamefont {L.~N.}\ \bibnamefont {Pfeiffer}}, \bibinfo {author}
  {\bibfnamefont {K.~W.}\ \bibnamefont {West}}, \bibinfo {author}
  {\bibfnamefont {K.~W.}\ \bibnamefont {Baldwin}},\ and\ \bibinfo {author}
  {\bibfnamefont {M.}~\bibnamefont {Shayegan}},\ }\bibfield  {title} {\bibinfo
  {title} {Spontaneous valley polarization of itinerant electrons},\ }\href
  {https://doi.org/10.1103/PhysRevLett.127.116601} {\bibfield  {journal}
  {\bibinfo  {journal} {Phys. Rev. Lett.}\ }\textbf {\bibinfo {volume} {127}},\
  \bibinfo {pages} {116601} (\bibinfo {year} {2021})}\BibitemShut {NoStop}%
\bibitem [{\citenamefont {Hossain}\ \emph {et~al.}(2022)\citenamefont
  {Hossain}, \citenamefont {Ma}, \citenamefont {Villegas-Rosales},
  \citenamefont {Chung}, \citenamefont {Pfeiffer}, \citenamefont {West},
  \citenamefont {Baldwin},\ and\ \citenamefont {Shayegan}}]{hossain2022}%
  \BibitemOpen
  \bibfield  {author} {\bibinfo {author} {\bibfnamefont {M.~S.}\ \bibnamefont
  {Hossain}}, \bibinfo {author} {\bibfnamefont {M.~K.}\ \bibnamefont {Ma}},
  \bibinfo {author} {\bibfnamefont {K.~A.}\ \bibnamefont {Villegas-Rosales}},
  \bibinfo {author} {\bibfnamefont {Y.~J.}\ \bibnamefont {Chung}}, \bibinfo
  {author} {\bibfnamefont {L.~N.}\ \bibnamefont {Pfeiffer}}, \bibinfo {author}
  {\bibfnamefont {K.~W.}\ \bibnamefont {West}}, \bibinfo {author}
  {\bibfnamefont {K.~W.}\ \bibnamefont {Baldwin}},\ and\ \bibinfo {author}
  {\bibfnamefont {M.}~\bibnamefont {Shayegan}},\ }\bibfield  {title} {\bibinfo
  {title} {Anisotropic two-dimensional disordered wigner solid},\ }\href
  {https://doi.org/10.1103/PhysRevLett.129.036601} {\bibfield  {journal}
  {\bibinfo  {journal} {Phys. Rev. Lett.}\ }\textbf {\bibinfo {volume} {129}},\
  \bibinfo {pages} {036601} (\bibinfo {year} {2022})}\BibitemShut {NoStop}%
\bibitem [{\citenamefont {Raines}\ \emph
  {et~al.}(2024{\natexlab{a}})\citenamefont {Raines}, \citenamefont {Glazman},\
  and\ \citenamefont {Chubukov}}]{raines2024unconventional}%
  \BibitemOpen
  \bibfield  {author} {\bibinfo {author} {\bibfnamefont {Z.~M.}\ \bibnamefont
  {Raines}}, \bibinfo {author} {\bibfnamefont {L.~I.}\ \bibnamefont
  {Glazman}},\ and\ \bibinfo {author} {\bibfnamefont {A.~V.}\ \bibnamefont
  {Chubukov}},\ }\bibfield  {title} {\bibinfo {title} {Unconventional
  discontinuous transitions in a two-dimensional system with spin and valley
  degrees of freedom},\ }\href {https://doi.org/10.1103/PhysRevB.110.155402}
  {\bibfield  {journal} {\bibinfo  {journal} {Phys. Rev. B}\ }\textbf {\bibinfo
  {volume} {110}},\ \bibinfo {pages} {155402} (\bibinfo {year}
  {2024}{\natexlab{a}})}\BibitemShut {NoStop}%
\bibitem [{\citenamefont {Raines}\ \emph
  {et~al.}(2024{\natexlab{b}})\citenamefont {Raines}, \citenamefont {Glazman},\
  and\ \citenamefont {Chubukov}}]{raines2024isospin}%
  \BibitemOpen
  \bibfield  {author} {\bibinfo {author} {\bibfnamefont {Z.~M.}\ \bibnamefont
  {Raines}}, \bibinfo {author} {\bibfnamefont {L.~I.}\ \bibnamefont
  {Glazman}},\ and\ \bibinfo {author} {\bibfnamefont {A.~V.}\ \bibnamefont
  {Chubukov}},\ }\bibfield  {title} {\bibinfo {title} {Unconventional
  discontinuous transitions in isospin systems},\ }\href
  {https://doi.org/10.1103/PhysRevLett.133.146501} {\bibfield  {journal}
  {\bibinfo  {journal} {Phys. Rev. Lett.}\ }\textbf {\bibinfo {volume} {133}},\
  \bibinfo {pages} {146501} (\bibinfo {year} {2024}{\natexlab{b}})}\BibitemShut
  {NoStop}%
\bibitem [{\citenamefont {Zhang}\ \emph {et~al.}(2010)\citenamefont {Zhang},
  \citenamefont {Sahu}, \citenamefont {Min},\ and\ \citenamefont
  {MacDonald}}]{zhang2010}%
  \BibitemOpen
  \bibfield  {author} {\bibinfo {author} {\bibfnamefont {F.}~\bibnamefont
  {Zhang}}, \bibinfo {author} {\bibfnamefont {B.}~\bibnamefont {Sahu}},
  \bibinfo {author} {\bibfnamefont {H.}~\bibnamefont {Min}},\ and\ \bibinfo
  {author} {\bibfnamefont {A.~H.}\ \bibnamefont {MacDonald}},\ }\bibfield
  {title} {\bibinfo {title} {Band structure of $abc$-stacked graphene
  trilayers},\ }\href {https://doi.org/10.1103/PhysRevB.82.035409} {\bibfield
  {journal} {\bibinfo  {journal} {Phys. Rev. B}\ }\textbf {\bibinfo {volume}
  {82}},\ \bibinfo {pages} {035409} (\bibinfo {year} {2010})}\BibitemShut
  {NoStop}%
\bibitem [{\citenamefont {Min}\ and\ \citenamefont
  {MacDonald}(2008)}]{min2008}%
  \BibitemOpen
  \bibfield  {author} {\bibinfo {author} {\bibfnamefont {H.}~\bibnamefont
  {Min}}\ and\ \bibinfo {author} {\bibfnamefont {A.~H.}\ \bibnamefont
  {MacDonald}},\ }\bibfield  {title} {\bibinfo {title} {Electronic structure of
  multilayer graphene},\ }\href {https://doi.org/10.1143/PTPS.176.227}
  {\bibfield  {journal} {\bibinfo  {journal} {Progress of Theoretical Physics
  Supplement}\ }\textbf {\bibinfo {volume} {176}},\ \bibinfo {pages} {227}
  (\bibinfo {year} {2008})},\ \Eprint
  {https://arxiv.org/abs/https://academic.oup.com/ptps/article-pdf/doi/10.1143/PTPS.176.227/5322668/176-227.pdf}
  {https://academic.oup.com/ptps/article-pdf/doi/10.1143/PTPS.176.227/5322668/176-227.pdf}
  \BibitemShut {NoStop}%
\bibitem [{\citenamefont {Zhang}\ \emph {et~al.}(2023)\citenamefont {Zhang},
  \citenamefont {Polski}, \citenamefont {Thomson}, \citenamefont
  {Lantagne-Hurtubise}, \citenamefont {Lewandowski}, \citenamefont {Zhou},
  \citenamefont {Watanabe}, \citenamefont {Taniguchi}, \citenamefont {Alicea},\
  and\ \citenamefont {Nadj-Perge}}]{zhang2023}%
  \BibitemOpen
  \bibfield  {author} {\bibinfo {author} {\bibfnamefont {Y.}~\bibnamefont
  {Zhang}}, \bibinfo {author} {\bibfnamefont {R.}~\bibnamefont {Polski}},
  \bibinfo {author} {\bibfnamefont {A.}~\bibnamefont {Thomson}}, \bibinfo
  {author} {\bibfnamefont {{\'E}.}~\bibnamefont {Lantagne-Hurtubise}}, \bibinfo
  {author} {\bibfnamefont {C.}~\bibnamefont {Lewandowski}}, \bibinfo {author}
  {\bibfnamefont {H.}~\bibnamefont {Zhou}}, \bibinfo {author} {\bibfnamefont
  {K.}~\bibnamefont {Watanabe}}, \bibinfo {author} {\bibfnamefont
  {T.}~\bibnamefont {Taniguchi}}, \bibinfo {author} {\bibfnamefont
  {J.}~\bibnamefont {Alicea}},\ and\ \bibinfo {author} {\bibfnamefont
  {S.}~\bibnamefont {Nadj-Perge}},\ }\bibfield  {title} {\bibinfo {title}
  {Enhanced superconductivity in spin--orbit proximitized bilayer graphene},\
  }\href {https://doi.org/10.1038/s41586-022-05446-x} {\bibfield  {journal}
  {\bibinfo  {journal} {Nature}\ }\textbf {\bibinfo {volume} {613}},\ \bibinfo
  {pages} {268} (\bibinfo {year} {2023})}\BibitemShut {NoStop}%
\bibitem [{\citenamefont {Patterson}\ \emph {et~al.}(2024)\citenamefont
  {Patterson}, \citenamefont {Sheekey}, \citenamefont {Arp}, \citenamefont
  {Holleis}, \citenamefont {Koh}, \citenamefont {Choi}, \citenamefont {Xie},
  \citenamefont {Xu}, \citenamefont {Redekop}, \citenamefont {Babikyan},
  \citenamefont {Zhou}, \citenamefont {Cheng}, \citenamefont {Taniguchi},
  \citenamefont {Watanabe}, \citenamefont {Jin}, \citenamefont
  {{Lantagne-Hurtubise}}, \citenamefont {Alicea},\ and\ \citenamefont
  {Young}}]{Patterson2024}%
  \BibitemOpen
  \bibfield  {author} {\bibinfo {author} {\bibfnamefont {C.~L.}\ \bibnamefont
  {Patterson}}, \bibinfo {author} {\bibfnamefont {O.~I.}\ \bibnamefont
  {Sheekey}}, \bibinfo {author} {\bibfnamefont {T.~B.}\ \bibnamefont {Arp}},
  \bibinfo {author} {\bibfnamefont {L.~F.~W.}\ \bibnamefont {Holleis}},
  \bibinfo {author} {\bibfnamefont {J.~M.}\ \bibnamefont {Koh}}, \bibinfo
  {author} {\bibfnamefont {Y.}~\bibnamefont {Choi}}, \bibinfo {author}
  {\bibfnamefont {T.}~\bibnamefont {Xie}}, \bibinfo {author} {\bibfnamefont
  {S.}~\bibnamefont {Xu}}, \bibinfo {author} {\bibfnamefont {E.}~\bibnamefont
  {Redekop}}, \bibinfo {author} {\bibfnamefont {G.}~\bibnamefont {Babikyan}},
  \bibinfo {author} {\bibfnamefont {H.}~\bibnamefont {Zhou}}, \bibinfo {author}
  {\bibfnamefont {X.}~\bibnamefont {Cheng}}, \bibinfo {author} {\bibfnamefont
  {T.}~\bibnamefont {Taniguchi}}, \bibinfo {author} {\bibfnamefont
  {K.}~\bibnamefont {Watanabe}}, \bibinfo {author} {\bibfnamefont
  {C.}~\bibnamefont {Jin}}, \bibinfo {author} {\bibfnamefont {E.}~\bibnamefont
  {{Lantagne-Hurtubise}}}, \bibinfo {author} {\bibfnamefont {J.}~\bibnamefont
  {Alicea}},\ and\ \bibinfo {author} {\bibfnamefont {A.~F.}\ \bibnamefont
  {Young}},\ }\href@noop {} {\bibinfo {title} {Superconductivity and spin
  canting in spin-orbit proximitized rhombohedral trilayer graphene}} (\bibinfo
  {year} {2024}),\ \Eprint {https://arxiv.org/abs/2408.10190} {arXiv:2408.10190
  [cond-mat.mes-hall]} \BibitemShut {NoStop}%
\bibitem [{Note1()}]{Note1}%
  \BibitemOpen
  \bibinfo {note} {There is an evidence that a half-metal state is
  spin-polarized~\cite {barrera2022,seiler2022}.}\BibitemShut {Stop}%
\bibitem [{\citenamefont {Chai}\ \emph {et~al.}(2020)\citenamefont {Chai},
  \citenamefont {Chaudhuri}, \citenamefont {Choi}, \citenamefont {Komargodski},
  \citenamefont {Rabinovici},\ and\ \citenamefont {Smolkin}}]{Chai2020}%
  \BibitemOpen
  \bibfield  {author} {\bibinfo {author} {\bibfnamefont {N.}~\bibnamefont
  {Chai}}, \bibinfo {author} {\bibfnamefont {S.}~\bibnamefont {Chaudhuri}},
  \bibinfo {author} {\bibfnamefont {C.}~\bibnamefont {Choi}}, \bibinfo {author}
  {\bibfnamefont {Z.}~\bibnamefont {Komargodski}}, \bibinfo {author}
  {\bibfnamefont {E.}~\bibnamefont {Rabinovici}},\ and\ \bibinfo {author}
  {\bibfnamefont {M.}~\bibnamefont {Smolkin}},\ }\bibfield  {title} {\bibinfo
  {title} {Thermal order in conformal theories},\ }\href
  {https://doi.org/10.1103/PhysRevD.102.065014} {\bibfield  {journal} {\bibinfo
   {journal} {Phys. Rev. D}\ }\textbf {\bibinfo {volume} {102}},\ \bibinfo
  {pages} {065014} (\bibinfo {year} {2020})}\BibitemShut {NoStop}%
\bibitem [{\citenamefont {Kanamori}(1963)}]{Kanamori1963}%
  \BibitemOpen
  \bibfield  {author} {\bibinfo {author} {\bibfnamefont {J.}~\bibnamefont
  {Kanamori}},\ }\bibfield  {title} {\bibinfo {title} {Electron {{Correlation}}
  and {{Ferromagnetism}} of {{Transition Metals}}},\ }\href
  {https://doi.org/10.1143/PTP.30.275} {\bibfield  {journal} {\bibinfo
  {journal} {Prog. Theor. Phys.}\ }\textbf {\bibinfo {volume} {30}},\ \bibinfo
  {pages} {275} (\bibinfo {year} {1963})}\BibitemShut {NoStop}%
\bibitem [{\citenamefont {Raines}\ and\ \citenamefont
  {Chubukov}(2025)}]{raines2025}%
  \BibitemOpen
  \bibfield  {author} {\bibinfo {author} {\bibfnamefont {Z.~M.}\ \bibnamefont
  {Raines}}\ and\ \bibinfo {author} {\bibfnamefont {A.~V.}\ \bibnamefont
  {Chubukov}},\ }\href {https://arxiv.org/abs/2507.00158} {\bibinfo {title}
  {Superconductivity via paramagnon and magnon exchange in a 2d
  near-ferromagnetic full metal and ferromagnetic half-metal}} (\bibinfo {year}
  {2025}),\ \Eprint {https://arxiv.org/abs/2507.00158} {arXiv:2507.00158
  [cond-mat.supr-con]} \BibitemShut {NoStop}%
\bibitem [{\citenamefont {Wood}(1992)}]{polylog}%
  \BibitemOpen
  \bibfield  {author} {\bibinfo {author} {\bibfnamefont {D.}~\bibnamefont
  {Wood}},\ }\href {http://www.cs.kent.ac.uk/pubs/1992/110} {\emph {\bibinfo
  {title} {The Computation of Polylogarithms}}},\ \bibinfo {type} {Tech. Rep.}\
  \bibinfo {number} {15-92*}\ (\bibinfo  {institution} {University of Kent,
  Computing Laboratory},\ \bibinfo {address} {University of Kent, Canterbury,
  UK},\ \bibinfo {year} {1992})\BibitemShut {NoStop}%
\end{thebibliography}%

\end{document}